\documentclass[aps,prd,twocolumn,superscriptaddress,nofootinbib,floatfix]{revtex4-2}

\usepackage{amsmath,amssymb}
\usepackage{graphicx}
\usepackage{booktabs}
\usepackage{xcolor}
\usepackage{slashed}
\usepackage[colorlinks=true,linkcolor=blue,citecolor=blue,urlcolor=blue]{hyperref}
\hypersetup{pdftitle={Mass reach for vector-like lepton doublets, singlets and triplets at FCC-hh with NLO+NNLL predictions},
  pdfauthor={Haluk Denizli, Aysu Cavus, Abdulkadir Senol}}

\graphicspath{{figs/}}

\newcommand{\TeV}{\,\mathrm{TeV}}
\newcommand{\GeV}{\,\mathrm{GeV}}
\newcommand{\abinv}{\,\mathrm{ab}^{-1}}

\newcommand{\sqs}{\sqrt{s}}
\newcommand{\MET}{E_T^{\rm miss}}

\newcommand{\taueh}{\tau_h}
\newcommand{\nlonnll}{NLO+NNLL}
\newcommand{\LT}{L_T}

\begin{document}

\title{Mass reach for vector-like lepton doublets, singlets and triplets\\
       at FCC-hh with NLO+NNLL predictions}

\author{H. Denizli}
\email{denizli\_h@ibu.edu.tr}
\affiliation{Department of Physics, Bolu Abant \.{I}zzet Baysal University, 14030 Bolu, T\"urkiye}
\author{A. Cavus}
\email{aysucavus21@gmail.com}
\affiliation{Department of Physics, Bolu Abant \.{I}zzet Baysal University, 14030 Bolu, T\"urkiye}
\author{A. Senol}
\email{senol\_a@ibu.edu.tr}
\affiliation{Department of Physics, Bolu Abant \.{I}zzet Baysal University, 14030 Bolu, T\"urkiye}

\date{\today}

\begin{abstract}
Vector-like leptons (VLLs) appear in many extensions of the Standard
Model, including composite Higgs, grand-unified and flavour models.
Their electroweak pair production makes them a natural target for a
future hadron collider. With the baseline energy of the Future Circular
hadron collider (FCC-hh) now set to $\sqs = 84.6\TeV$, we assess the
reach for VLLs at this energy and at $100\TeV$. Drell--Yan pair
production is computed at next-to-leading order in QCD matched to
threshold resummation at next-to-next-to-leading logarithmic accuracy
(\nlonnll), including scale and parton-distribution uncertainties.
These predictions are combined with detector-level searches for
doublet, singlet and triplet VLLs mixing with taus, electrons, muons or
any combination of them. For $30\abinv$ at $84.6\TeV$ and a 10\%
background uncertainty, multilepton final states exclude doublets at
95\% CL up to $6.4\TeV$ for tau mixing and up to $9.3$--$9.65\TeV$ for
electron or muon mixing. Singlets seldom produce several charged
leptons, but a search for an opposite-sign lepton pair and a boosted
$Z$- or Higgs-boson jet excludes them up to $3.45\TeV$ for tau mixing, a scenario for which
earlier projections found no sensitivity, and $5.7$--$6.0\TeV$ for
electron or muon mixing. Type-III triplets,
whose Majorana neutral states give same-sign dileptons, are excluded up
to $5.6\TeV$ and $8.4$--$8.75\TeV$ in the same two cases. At $100\TeV$
the reach for electron and muon mixing is 11--16\% higher. Compared
with a leading-order normalisation, the \nlonnll{} predictions raise
the reach by $0.25$--$0.5\TeV$ and reduce its scale uncertainty from up
to $170\GeV$ to below $5\GeV$. The remaining theoretical uncertainty of
$60$--$200\GeV$ comes almost entirely from parton distributions at
large momentum fractions.
\end{abstract}

\maketitle

%% ===================================================================
\section{Introduction}
\label{sec:intro}
%% ===================================================================

Vector-like fermions, whose left- and right-handed components transform
identically under the Standard Model (SM) gauge group, admit
gauge-invariant mass terms that do not originate from electroweak
symmetry breaking. Unlike a fourth chiral generation, they are
anomaly-free on their own and, because their mass does not arise from
the Higgs mechanism, they decouple smoothly and are not excluded by
Higgs-production rates or electroweak precision
data~\cite{Eberhardt:2012gv}. Vector-like leptons (VLLs) appear in
composite Higgs models~\cite{Agashe:2004rs,Redi:2013pga}, in
grand-unified theories such as $E_6$~\cite{Hewett:1988xc}, and, as
fermion triplets, in the type-III seesaw mechanism for neutrino
masses~\cite{Foot:1988aq}. Independently of any specific model, VLLs are
among the few fields that can couple renormalisably to the SM leptons
and the Higgs doublet, and they arise as tree-level UV completions of
dimension-six operators involving leptons~\cite{deBlas:2017xtg}. Their mixing with the SM
leptons modifies the $Z$ and $W$ couplings, the Higgs decays and the
anomalous magnetic moments, and it induces lepton-flavour-violating
processes. For this reason VLLs are a common ingredient in explanations of
flavour and precision anomalies~\cite{Poh:2017tfo,Crivellin:2020ebi}.
These indirect effects are suppressed by the mixing angles, whereas pair
production through gauge interactions depends only on the mass and the
SU(2)$\times$U(1) quantum numbers. Direct searches at hadron colliders
therefore provide a model-independent probe, complementary to precision
measurements. The indirect and collider phenomenology of VLLs has been
studied
extensively~\cite{delAguila:2008pw,Falkowski:2013jya,Kumar:2015tna,Crivellin:2020ebi}.

At the LHC, multilepton searches exclude the tau-coupled doublet below
$1045\GeV$ (CMS~\cite{CMS:2022nty}) and in the range $130$--$900\GeV$
(ATLAS~\cite{ATLAS:2023sbu}, see also Ref.~\cite{CMS:2019hsm}), and the
tau-coupled singlet only between $125$ and $150\GeV$~\cite{CMS:2022nty}.
ATLAS has excluded doublet vector-like electrons and muons below
$1220\GeV$ and $1270\GeV$, respectively, and singlets below
$320$--$400\GeV$~\cite{ATLAS:2024mrr}. Type-III seesaw triplets are excluded
below $0.85$--$1.07\TeV$, depending on the flavour
structure~\cite{CMS:2022nty}, and below $0.91\TeV$ for equal branching
fractions~\cite{ATLAS:2022yhd}. The CMS searches are reviewed in
Ref.~\cite{CMS:2024bni}.

The FCC feasibility study~\cite{FCC:2025uan} has updated the FCC-hh baseline to a centre-of-mass energy of about
$85\TeV$ ($\sqs = 84.6\TeV$) instead of the $100\TeV$ assumed in earlier
studies~\cite{FCC:2018vvp}. To our knowledge, the only dedicated FCC-hh projection for VLL pair
production~\cite{Bhattiprolu:2019vdu} predates this change and considers
VLLs coupling exclusively to taus. Using leading-order cross sections
and a cut-and-count analysis, it finds a reach of about $5.75\TeV$ for
the tau-coupled doublet at $100\TeV$ with $30\abinv$ and no sensitivity
to the minimal singlet. For singlets mixing with electrons or muons, an
extrapolation of LHC searches to a $100\TeV$ collider with $3\abinv$
gives a reach of about $2.5\TeV$~\cite{Guedes:2021oqx}, and chains of
such singlets motivated by low-scale flavour models have recently been
studied in the same context~\cite{Han:2026ygh}.

The present work extends the study of Ref.~\cite{Bhattiprolu:2019vdu}
in three respects.
(i)~Building on the LHC calculation of Ref.~\cite{AH:2023hft}, we
provide \nlonnll{} predictions for Drell--Yan pair production of
doublet, singlet and triplet VLLs at $84.6$ and $100\TeV$, including scale and PDF
uncertainties and differential distributions.
(ii)~We extend the analysis from the tau-coupled doublet to doublet,
singlet and triplet VLLs mixing with $\tau$, $e$ and $\mu$, and optimise
multilepton signal regions for all flavours. Since the singlet
rarely yields multilepton final states, we design dedicated searches
for it, based on an opposite-sign lepton or hadronic-tau pair and a
boosted $Z/h$ jet. Results are given for pure-flavour benchmarks and
across the plane of flavour branching fractions at both $84.6$ and
$100\TeV$.
(iii)~We quantify how the precision of the signal prediction affects
the projected mass reach for all three representations at both
energies.

The paper is organised as follows. After introducing the models in
Sec.~\ref{sec:models}, we present the precision predictions in
Sec.~\ref{sec:xsec}. The simulation setup and the analysis strategy are
described in Secs.~\ref{sec:sim} and~\ref{sec:analysis}, and our
results are given in Sec.~\ref{sec:results}. We then compare the FCC-hh
reach with that of other facilities in Sec.~\ref{sec:compl} before
concluding in Sec.~\ref{sec:concl}. Appendix~\ref{app:lhc} validates
our calculation against the LHC results of Ref.~\cite{AH:2023hft}.

%% ===================================================================
\section{Models}
\label{sec:models}
%% ===================================================================

We consider the minimal scenarios of
Refs.~\cite{Kumar:2015tna,Bhattiprolu:2019vdu,AH:2023hft}, namely a
doublet $L_{L,R} = (N, E)^T \sim (\mathbf{1},\mathbf{2},-1/2)$, a
singlet $\tilde E_{L,R} \sim (\mathbf{1},\mathbf{1},-1)$ and a
type-III seesaw triplet~\cite{Foot:1988aq}
$\Sigma \sim (\mathbf{1},\mathbf{3},0)$ with charged and neutral
(Majorana) components $\Sigma^\pm$ and $\Sigma^0$. Each multiplet mixes
through Yukawa couplings with the SM leptons of flavour
$\ell = e,\mu,\tau$,
\begin{align}
  \mathcal{L}_{\rm D} &\supset - M_L\, \bar{L}_L L_R
  - \sum_{\ell} \lambda_\ell\, \bar{L}_L H \ell_R + \text{h.c.}, \\
  \mathcal{L}_{\rm S} &\supset - M_L\, \bar{\tilde E}_L \tilde E_R
  - \sum_{\ell} \lambda_\ell\, \bar{l}_{\ell L} H \tilde E_R + \text{h.c.},
\end{align}
where $l_{\ell L}$ and $\ell_R$ denote the SM lepton doublets and
singlets and $H$ is the Higgs doublet. For the triplet we use the
type-III seesaw Lagrangian implemented in Ref.~\cite{AH:2023hft}.
After electroweak symmetry breaking with $\langle H\rangle = v/\sqrt{2}$,
the charged-lepton mass matrix in the basis $(\ell, E)$ acquires the
off-diagonal entry $\lambda_\ell v/\sqrt{2}$ that couples $\ell_R$ to
$L_L$ for the doublet and $l_{\ell L}$ to $\tilde E_R$ for the singlet.
For $M_L \gg v$ the heavy state mixes with the SM lepton through the
angle
\begin{equation}
  \epsilon_\ell \simeq \frac{\lambda_\ell v}{\sqrt{2} M_L}
\end{equation}
that describes right-handed (left-handed) mixing for the doublet
(singlet), while the mixing of the other chirality is suppressed by
$m_\ell/M_L$~\cite{Kumar:2015tna}. The neutral doublet component $N$
does not mix at this order, because the SM contains no right-handed
neutrino. The pair-production cross sections of Sec.~\ref{sec:xsec}
depend only on the gauge quantum numbers. We compute them for all three
multiplets and apply the detector-level analysis of
Secs.~\ref{sec:sim}--\ref{sec:results} to each of them.

We parametrise the flavour structure by the branching fractions
\begin{equation}
  \mathcal{B}_e + \mathcal{B}_\mu + \mathcal{B}_\tau = 1,
  \qquad \mathcal{B}_\ell \propto |\epsilon_\ell|^2 ,
\end{equation}
and assume that the mixing is small enough to evade indirect bounds.
Electroweak precision data constrain the mixing with a single flavour
to $|\epsilon_e| < 0.018$, $|\epsilon_\mu| < 0.034$ ($0.045$) and
$|\epsilon_\tau| < 0.030$ at 90\% CL for the singlet
(doublet)~\cite{delAguila:2008pw}. An updated global fit is given in
Ref.~\cite{Crivellin:2020ebi}. If the VLL mixes with more than one
flavour, the modified $Z$ couplings induce $\mu\to 3e$, $\mu$--$e$
conversion in nuclei and $\tau\to 3\ell$ at tree level, and
$\mu\to e\gamma$ and $\tau\to\ell\gamma$ at one
loop~\cite{Poh:2017tfo,Crivellin:2020ebi,MEGII:2025gzr}.
These processes constrain products of the mixings much more strongly
than electroweak precision data. With the tree-level expressions of
Ref.~\cite{Crivellin:2020ebi}, the bounds
$\mathcal{B}(\mu\to 3e) < 1.0\times10^{-12}$ and
$\mathcal{B}(\tau\to 3\mu) < 1.1\times10^{-8}$ translate into
$|\epsilon_e\epsilon_\mu| \lesssim 2\times10^{-6}$ and
$|\epsilon_\tau\epsilon_\mu| \lesssim 6\times10^{-4}$ for the doublet
and the singlet. Bounds of similar size apply to the
triplet~\cite{Crivellin:2020ebi}. For the $e$--$\mu$ products,
$\mu$--$e$ conversion is comparably or more restrictive. These limits
apply to a single VLL generation mixing with several flavours, as
considered here, and can be avoided if each SM flavour mixes with a
different VLL generation~\cite{Crivellin:2020ebi}. They do not,
however, restrict the collider signal, since the pair-production cross
section is independent of $\epsilon_\ell$. The total width scales as
$\Gamma \propto |\epsilon|^2 M_L^3/v^2$, so the decays are prompt for
$|\epsilon| \gtrsim 10^{-7}$ at $M_L \gtrsim 1\TeV$, whereas the
branching fractions depend only on the ratios
$|\epsilon_e|^2 : |\epsilon_\mu|^2 : |\epsilon_\tau|^2$. Mixed-flavour
scenarios with an overall mixing in the range
$10^{-7} \lesssim |\epsilon| \lesssim 10^{-4}$ are therefore
compatible with the LFV bounds and lead to the same collider
phenomenology.

\begin{figure}[t]
  \centering
  \includegraphics[width=\columnwidth]{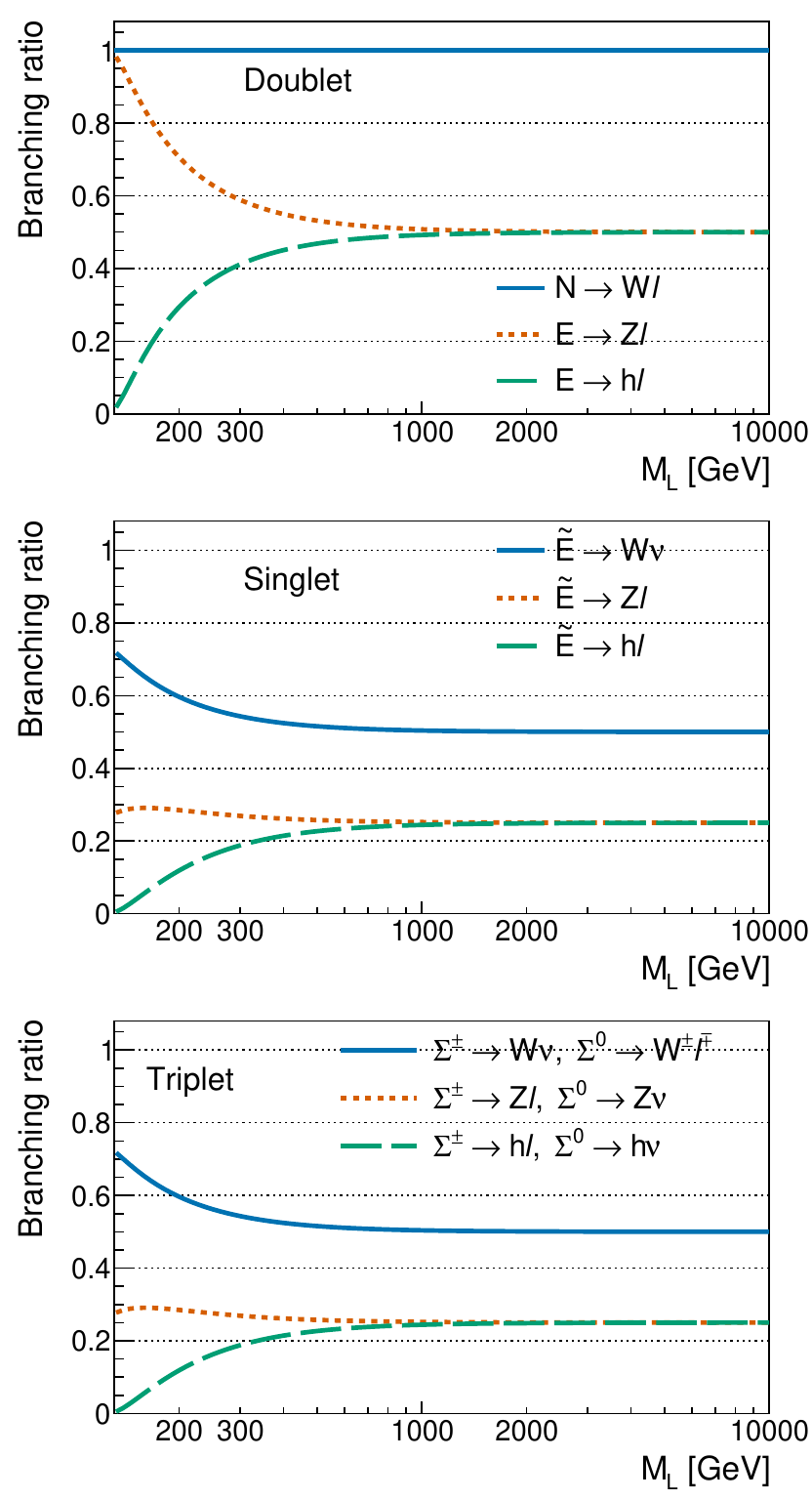}
  \caption{Branching ratios of the VLLs as a function of $M_L$ at
  leading order in the mixing ($\ell = e, \mu, \tau$), for the doublet
  (top), singlet (middle) and triplet (bottom). The charged and neutral
  triplet components have identical branching ratios.}
  \label{fig:br}
\end{figure}

In the heavy-mass limit the decays
$E^\pm \to W^\pm\nu, Z\ell^\pm, h\ell^\pm$ and
$N \to W^\pm\ell^\mp$ are governed by the Goldstone equivalence theorem. For the doublet, $E \to W\nu$ is suppressed, so
that $\mathcal{B}(E\to Z\ell) = \mathcal{B}(E\to h\ell) = 1/2$ and
$\mathcal{B}(N\to W\ell) = 1$~\cite{Kumar:2015tna,Bhattiprolu:2019vdu}.
For the singlet,
$\mathcal{B}(W):\mathcal{B}(Z):\mathcal{B}(h)$ approaches $2:1:1$, and
the same pattern holds for both triplet components,
$\Sigma^\pm \to W^\pm\nu, Z\ell^\pm, h\ell^\pm$ and
$\Sigma^0 \to W^\pm\ell^\mp, Z\nu, h\nu$. Since $\Sigma^0$ is a
Majorana fermion, it decays with equal probability to $W^+\ell^-$ and
$W^-\ell^+$. The models are implemented with the
FeynRules/UFO libraries of Ref.~\cite{AH:2023hft}, which include the
UV and $R_2$ counterterms required for NLO QCD calculations in
MadGraph5\_aMC@NLO~3.5.7~\cite{Alwall:2014hca}. Using these
implementations and the automatic width computation of
MadGraph5\_aMC@NLO, we have confirmed the triplet branching fractions at
$M_\Sigma = 5\TeV$.
At leading order in the mixing the partial widths
are
\begin{align}
  \Gamma(F\to V\ell) &\propto c_V\,(1-r_V)^2(1+2r_V), \nonumber\\
  \Gamma(F\to h\ell) &\propto c_h\,(1-r_h)^2, \qquad r_X = m_X^2/M_L^2 ,
\end{align}
with $(c_W, c_Z, c_h) = (0,1,1)$ for the doublet $E$ and $(2,1,1)$ for
the singlet~\cite{Kumar:2015tna} and for both triplet components. The
doublet $N$ decays only to $W\ell$. Figure~\ref{fig:br} shows the
resulting branching ratios. For $M_L \gtrsim 1\TeV$ they lie within one percentage point of their asymptotic values, so that at the multi-TeV
masses relevant for FCC-hh they depend neither on $M_L$ nor on the overall
size of the mixing.

%% ===================================================================
\section{Precision predictions at FCC-hh}
\label{sec:xsec}
%% ===================================================================

We consider pair production through the neutral-current (NC) and
charged-current (CC) Drell--Yan processes
\begin{equation}
  \begin{array}{l@{\quad}r@{\;\to\;}c@{\;\to\;}l@{\quad}l}
    \text{doublet} & q\bar q  & Z^*/\gamma^* & E^+E^-,\; N\bar N        & \text{(NC)} \\
                   & q\bar q' & W^{\pm *}    & E^\pm N                  & \text{(CC)} \\[3pt]
    \text{singlet} & q\bar q  & Z^*/\gamma^* & \tilde E^+\tilde E^-     & \text{(NC)} \\[3pt]
    \text{triplet} & q\bar q  & Z^*/\gamma^* & \Sigma^+\Sigma^-         & \text{(NC)} \\
                   & q\bar q' & W^{\pm *}    & \Sigma^\pm\Sigma^0       & \text{(CC)}
  \end{array}
  \label{eq:dy}
\end{equation}
The singlet has no neutral partner and is therefore produced only
through the neutral current.

\begin{figure*}[t]
  \centering
  \includegraphics[width=0.48\textwidth]{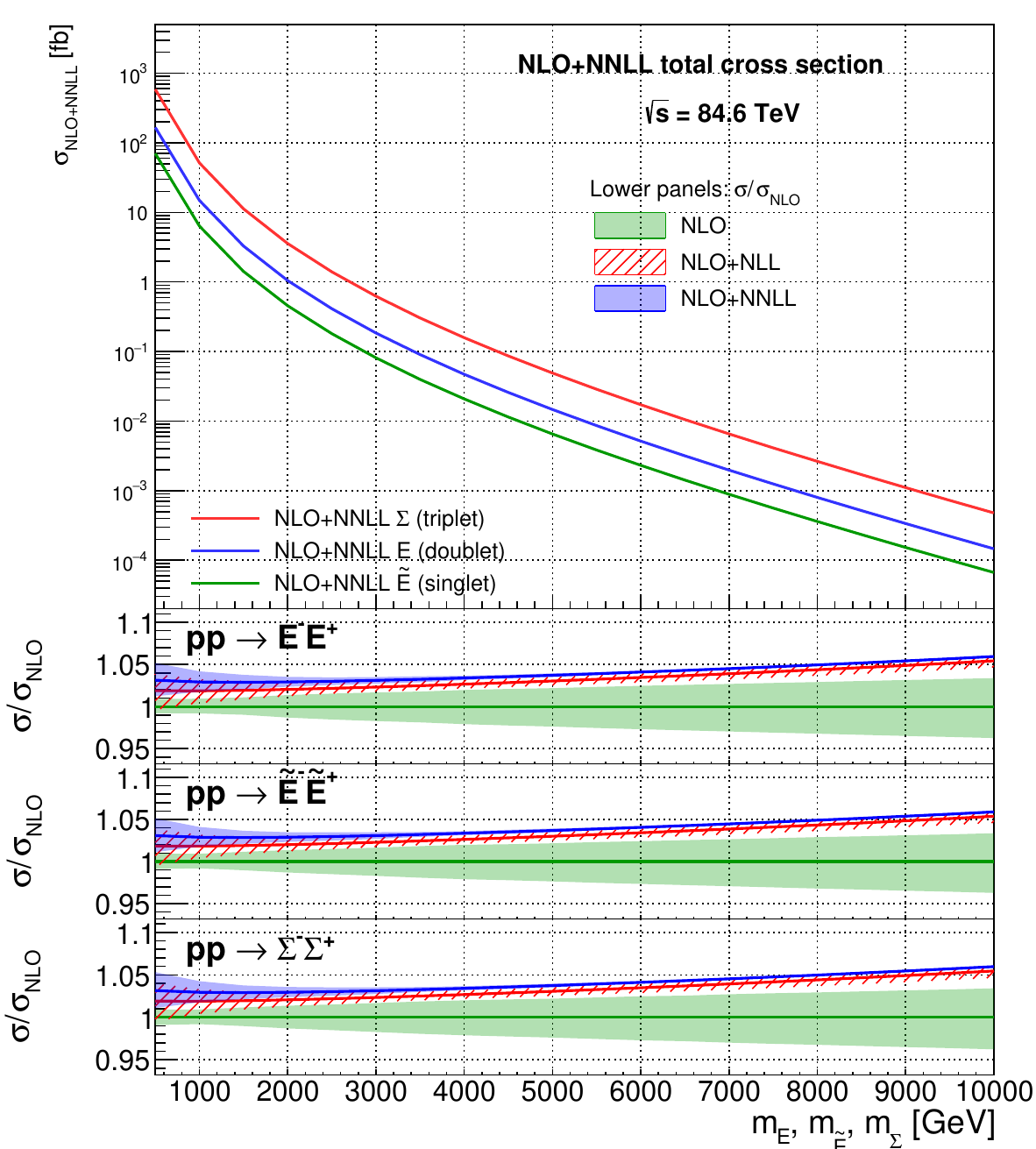}\hfill
  \includegraphics[width=0.48\textwidth]{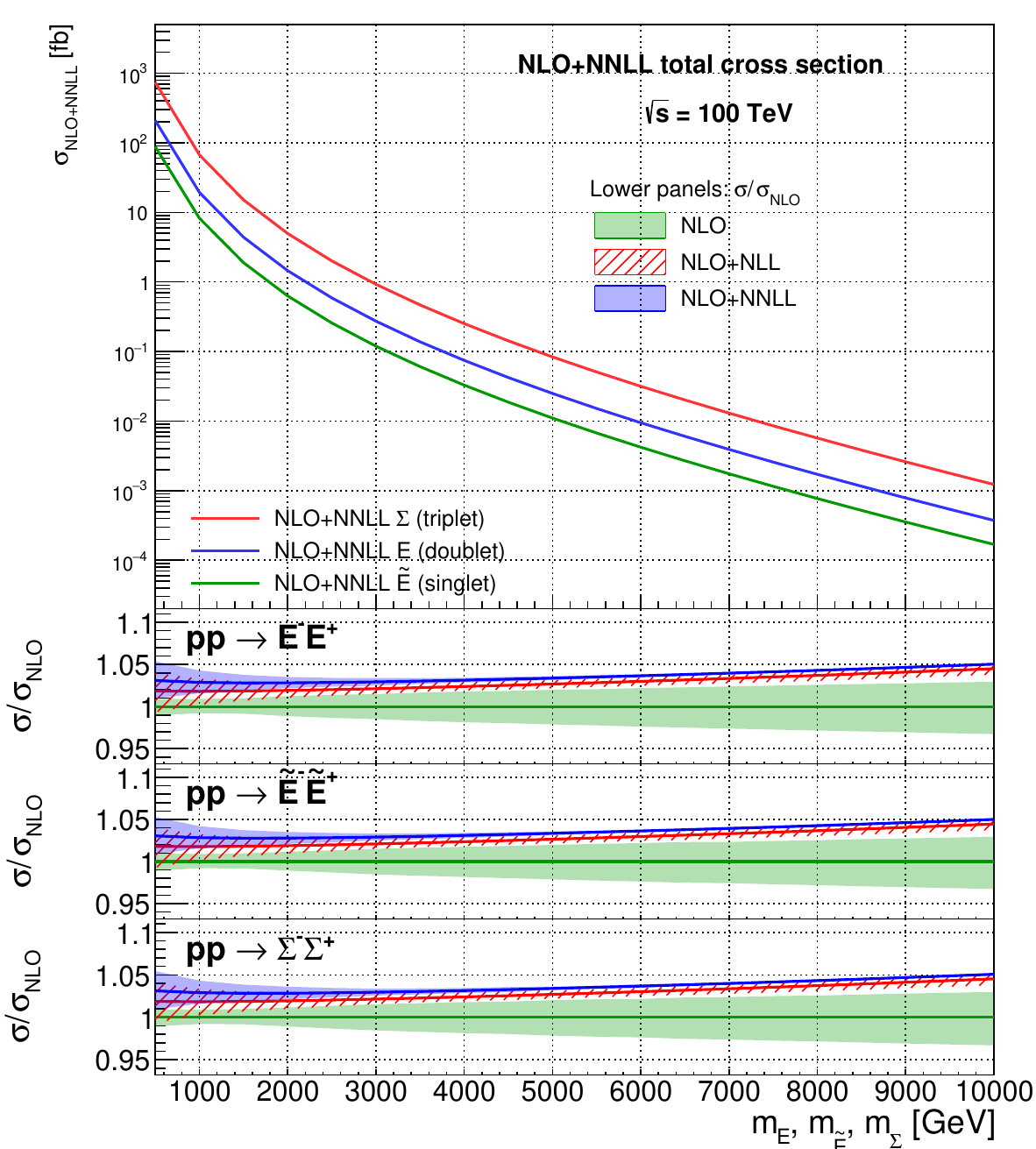}
  \caption{Total \nlonnll{} pair-production cross sections for the
  doublet ($E^+E^-$), singlet ($\tilde E^+\tilde E^-$) and triplet
  ($\Sigma^+\Sigma^-$) at $\sqs = 84.6\TeV$ (left) and $100\TeV$
  (right). The lower panels show the ratio to NLO together with the
  seven-point scale bands at NLO, NLO+NLL and \nlonnll.}
  \label{fig:xsec}
\end{figure*}

We compute these processes at NLO in QCD with
MadGraph5\_aMC@NLO in the five-flavour
scheme, using CT18NNLO~\cite{Hou:2019efy} parton densities through
LHAPDF~6~\cite{Buckley:2014ana}. The top-quark mass is set to
$172.7\GeV$ and all widths to zero. The central renormalisation and
factorisation scales are chosen as the invariant mass of the VLL pair, $\mu_R = \mu_F = \mu_0 = M_{LL}$.
Scale uncertainties are estimated from the envelope of the seven-point
variation $\mu_{R,F} \in \{\mu_0/2, \mu_0, 2\mu_0\}$, excluding the two
extreme combinations.

The NLO results are matched to threshold
resummation~\cite{Sterman:1986aj,Catani:1989ne} at NNLL accuracy in
Mellin space, where the convolution with the PDFs becomes a product and
the threshold logarithms $\ln(1-z)$, with $z = Q^2/\hat s$, become
$\ln N$,
\begin{multline}
  \sigma^{\text{\nlonnll}} = \sigma^{\rm NLO}
  + \int \! dQ^2\, \frac{\hat\sigma_0(Q^2)}{S}
  \oint \frac{dN}{2\pi i}\, \tau^{-N} \\
  \times f_{q}(N) f_{\bar q}(N)\, \Big[\Delta^{\rm res}_N - \Delta^{\rm res}_N\big|_{\mathcal O(\alpha_s)}\Big],
\end{multline}
where $\tau = Q^2/S$. The function $\Delta^{\rm res}_N$ contains the Drell--Yan
Sudakov exponent with the $g_1$, $g_2$ and $g_3$ functions of
Ref.~\cite{Catani:2003zt} in the $\bar N$ scheme, together with the
$N$-independent matching coefficients of Ref.~\cite{AH:2023hft}. Only
the $q\bar q$ channel is resummed. The inverse Mellin transform is
performed along a straight contour that lies to the right of the
singularities of the PDF moments and to the left of the Landau pole
(minimal prescription~\cite{Catani:1996yz}), using numerical Mellin
moments of the PDFs. We use our own implementation of this formalism,
which we have validated against
Ref.~\cite{AH:2023hft} at the LHC (App.~\ref{app:lhc}).

At multi-TeV masses the relevant momentum fractions reach
$x \sim 2M_L/\sqs \gtrsim 0.1$, where PDF uncertainties exceed the
residual scale dependence. We evaluate them at NLO with three modern
sets, CT18NNLO, MSHT20qed~\cite{Cridge:2021pxm} and
NNPDF4.0~QED~\cite{NNPDF:2024djq}. The PDF uncertainty is taken as the
envelope of their 68\% CL bands around the CT18NNLO central value,
with the CT18NNLO band rescaled from its default 90\% CL. This envelope
is applied as a relative uncertainty to
the \nlonnll{} prediction and added in quadrature to the scale
uncertainty.

Figure~\ref{fig:xsec} shows the \nlonnll{} cross sections for
$M_L = 0.5$--$10\TeV$ at $\sqs = 84.6$ and $100\TeV$. The NLO $K$ factor,
$\sigma^{\rm NLO}/\sigma^{\rm LO}$, increases from $1.13$ at $0.5\TeV$
to $1.29$ ($1.26$) at $10\TeV$ for $\sqs = 84.6\TeV$ ($100\TeV$),
independently of the multiplet. Threshold resummation adds a further
3\% for $M_L \lesssim 3\TeV$. Above this mass the correction grows,
reaching $3.4\%$, $4.1\%$ and $5.9\%$ at $4$, $6$ and $10\TeV$ for
$\sqs = 84.6\TeV$, whereas for $\sqs = 100\TeV$ it is slightly smaller,
reaching $5.0\%$ at $10\TeV$. The scale uncertainty decreases from $\pm(1$--$3.5)\%$ at
NLO to below $\pm 0.6\%$ ($\pm 0.8\%$) at \nlonnll{} for
$M_L \geq 2\TeV$ at $84.6\TeV$ ($100\TeV$). The PDF
uncertainty grows with mass, and for $E^+E^-$ production amounts to
${}^{+2.4}_{-3.0}\%$, ${}^{+6.3}_{-5.6}\%$ and ${}^{+11.7}_{-9.2}\%$ at
$2$, $6$ and $10\TeV$ for $\sqs = 84.6\TeV$, and to
${}^{+2.1}_{-2.7}\%$, ${}^{+5.2}_{-4.9}\%$ and ${}^{+9.4}_{-7.7}\%$ at
$100\TeV$. The envelopes for the singlet and the triplet agree with
that of $E^+E^-$ to within $0.5\%$ up to $6\TeV$ and to within $1\%$ at
8 and $10\TeV$, and that of $N\bar N$ agrees to within $0.7\%$ at all
masses. For the charged-current channel $E^\pm N$, and hence for
$\Sigma^\pm\Sigma^0$, the envelope is asymmetric, ${}^{+3.6}_{-3.1}\%$
at $2\TeV$ and ${}^{+10.0}_{-6.7}\%$ at $7\TeV$ for $\sqs = 84.6\TeV$,
because the NNPDF4.0 central value lies about 5\% above CT18NNLO. The
$E^\pm N$ and $N\bar N$ envelopes are computed up to $7\TeV$; at higher
masses we take their values at $7\TeV$ and scale them with the mass
dependence of the $E^+E^-$ envelope, which gives
${}^{+15.8}_{-9.8}\%$ for $E^\pm N$ at $10\TeV$.
At the masses relevant for the FCC-hh reach, the theoretical
uncertainty of the signal normalisation is therefore dominated entirely
by the PDFs. Table~\ref{tab:xsec} lists the \nlonnll{} cross sections
of the charged pairs $E^+E^-$, $\tilde E^+\tilde E^-$ and
$\Sigma^+\Sigma^-$, and Table~\ref{tab:xsec_cc} those of the
$E^\pm N$, $N\bar N$ and $\Sigma^\pm\Sigma^0$ channels. Together they
give the signal normalisation used in
Secs.~\ref{sec:sim}--\ref{sec:results}.

\begin{figure*}[t]
  \centering
  \includegraphics[width=0.32\textwidth]{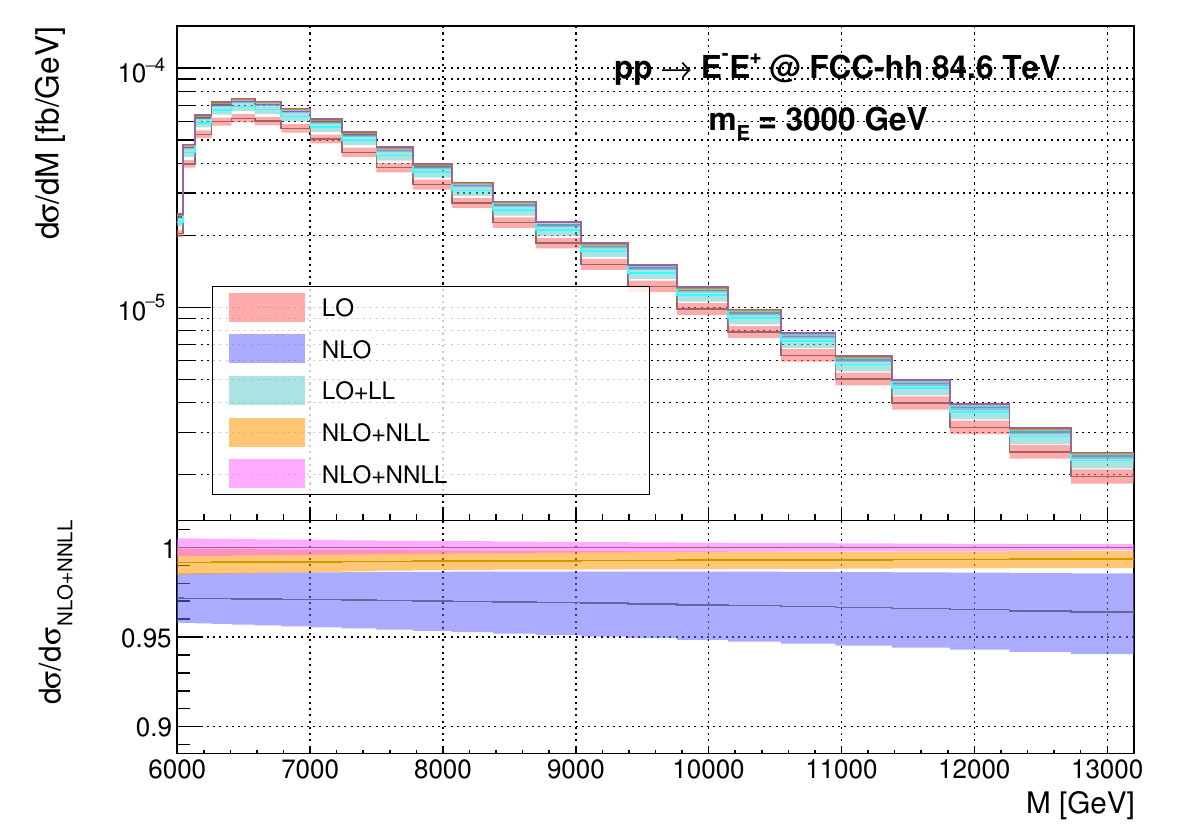}\hfill
  \includegraphics[width=0.32\textwidth]{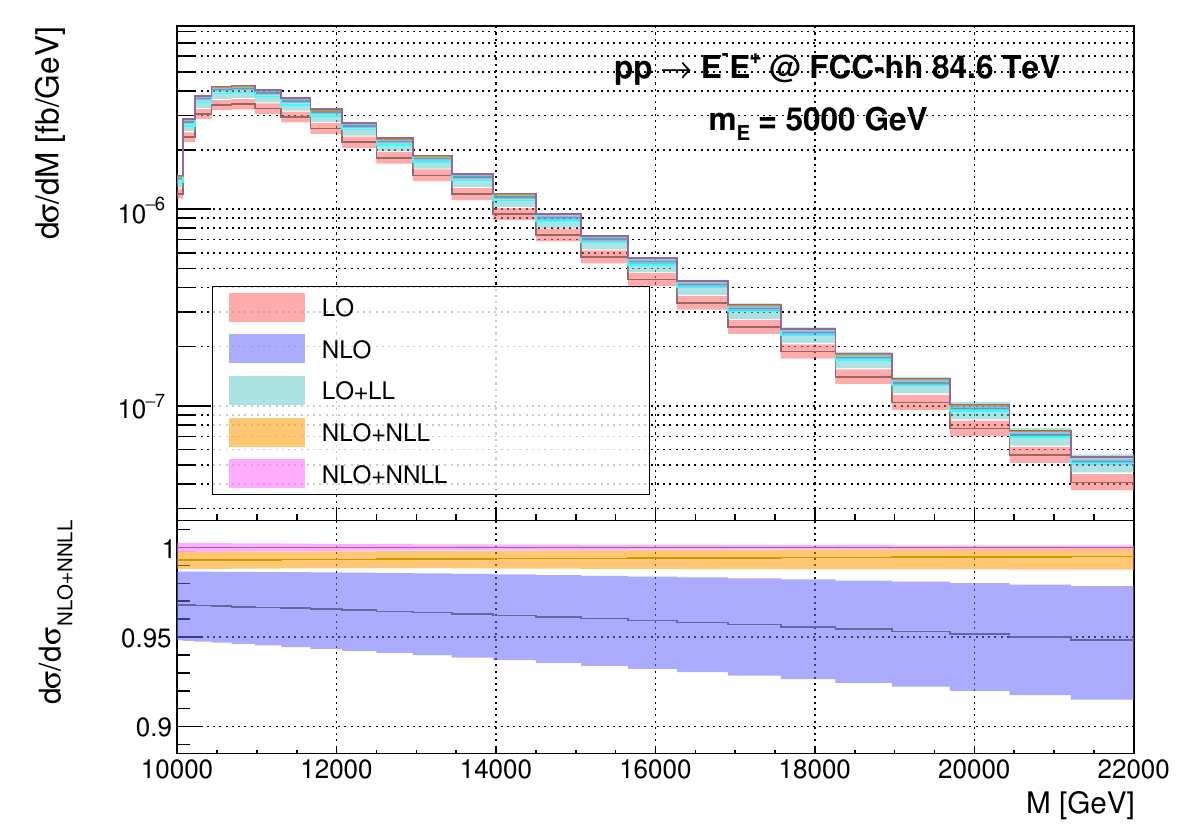}\hfill
  \includegraphics[width=0.32\textwidth]{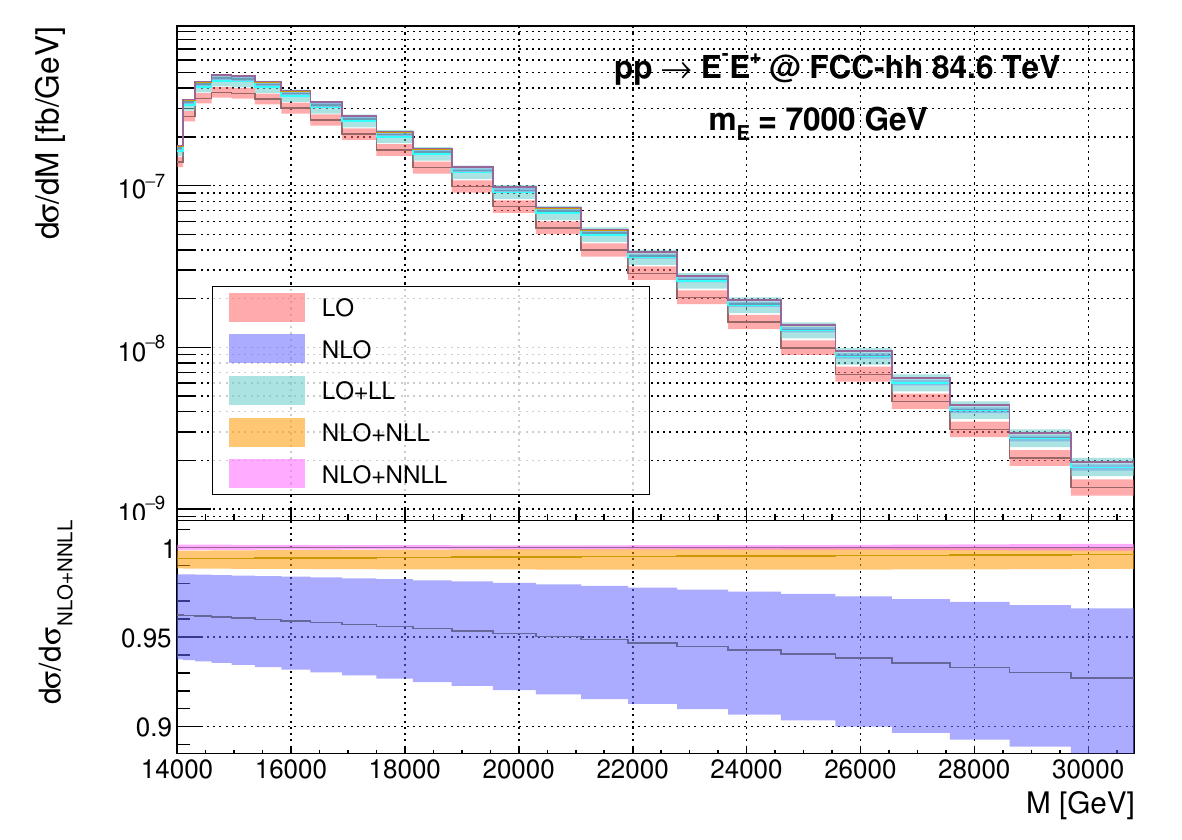}
  \caption{Invariant-mass distribution of the $E^+E^-$ pair at
  $\sqs = 84.6\TeV$ for $M_E = 3$, $5$ and $7\TeV$, at LO, NLO, LO+LL,
  NLO+NLL and \nlonnll, each with its seven-point scale band. The lower
  panels show the ratio of NLO, NLO+NLL and \nlonnll{} to \nlonnll.}
  \label{fig:dsdm}
\end{figure*}

Figure~\ref{fig:dsdm} shows the invariant-mass distribution of the
$E^+E^-$ pair at $84.6\TeV$ for $M_E = 3$, $5$ and $7\TeV$, from
threshold ($M_{LL} = 2M_E$) up to $M_{LL} = 4.4\,M_E$. The resummation
corrections grow towards large $M_{LL}$, so that the ratio NLO/\nlonnll{}
decreases from $0.96$--$0.97$ at threshold to $0.96$, $0.95$ and
$0.93$ at $M_{LL} = 4.4\,M_E$ for $M_E = 3$, $5$ and $7\TeV$,
respectively. The LO distribution lies 17--31\% below \nlonnll, and
the seven-point scale uncertainty decreases from $\pm(1.4$--$4.4)\%$
at NLO to below $\pm 0.5\%$ at \nlonnll. For the singlet
$\tilde E^+\tilde E^-$ and triplet $\Sigma^+\Sigma^-$ pairs, the ratios
to \nlonnll{} agree with those of the doublet to within $0.1\%$, so that
the distributions differ only in normalisation.

The transverse momentum of the VLL pair is generated by QCD radiation
and is therefore sensitive to the parton shower. Matching the NLO
calculation to Pythia~8.316~\cite{Bierlich:2022pfr} in the MC@NLO
formalism~\cite{Frixione:2002ik} with MC@NLO-$\Delta$
matching~\cite{Frederix:2020trv}, we find that NLO+PS exceeds the
fixed-order NLO prediction by a factor $1.55$--$1.8$ at
$p_T \approx 0.2$--$1\,M_L$, whereas the integrated cross sections agree
to better than $0.5\%$. With standard MC@NLO matching, the maximum is
$2.1$--$2.3$ and varies between $1.6$ and $2.7$ when the shower starting
scale is rescaled by factors of $2$ and $1/2$, a matching uncertainty
similar to that reported for other
processes~\cite{Heinrich:2017kxx,Jager:2020hkz}. The singlet and
triplet pairs behave in the same way, since the shape is determined by
initial-state radiation. These effects matter for analyses that rely on
the recoil of the pair but not for the inclusive rates used below.

\begin{table*}[t]
  \caption{\nlonnll{} cross sections (in fb) for charged VLL pair
  production at FCC-hh, with $\mu_0 = M_{LL}$ and CT18NNLO. The upper and
  lower values give the seven-point scale uncertainties in \%. The PDF
  uncertainties are discussed in Sec.~\ref{sec:xsec}.}
  \label{tab:xsec}
  \begin{ruledtabular}
  \begin{tabular}{c ccc ccc}
    & \multicolumn{3}{c}{$\sqs = 84.6\TeV$} & \multicolumn{3}{c}{$\sqs = 100\TeV$} \\
    $M_L$ [TeV] & $E^+E^-$ & $\tilde E^+\tilde E^-$ & $\Sigma^+\Sigma^-$
                & $E^+E^-$ & $\tilde E^+\tilde E^-$ & $\Sigma^+\Sigma^-$ \\
    \hline
    1  & $14.88^{+1.2}_{-1.1}$ & $6.369^{+1.2}_{-1.1}$ & $51.48^{+1.3}_{-1.1}$
       & $19.36^{+1.4}_{-1.3}$ & $8.254^{+1.3}_{-1.2}$ & $67.20^{+1.4}_{-1.3}$ \\
    2  & $1.047^{+0.6}_{-0.5}$ & $0.4554^{+0.6}_{-0.5}$ & $3.569^{+0.6}_{-0.6}$
       & $1.454^{+0.7}_{-0.6}$ & $0.6303^{+0.7}_{-0.7}$ & $4.975^{+0.8}_{-0.7}$ \\
    4  & $4.705\times10^{-2}\,{}^{+0.3}_{-0.2}$ & $2.085\times10^{-2}\,{}^{+0.2}_{-0.2}$ & $0.1581^{+0.2}_{-0.2}$
       & $7.462\times10^{-2}\,{}^{+0.3}_{-0.2}$ & $3.287\times10^{-2}\,{}^{+0.3}_{-0.3}$ & $0.2518^{+0.3}_{-0.3}$ \\
    6  & $5.158\times10^{-3}\,{}^{+0.1}_{-0.1}$ & $2.314\times10^{-3}\,{}^{+0.2}_{-0.2}$ & $1.721\times10^{-2}\,{}^{+0.1}_{-0.1}$
       & $9.468\times10^{-3}\,{}^{+0.2}_{-0.1}$ & $4.213\times10^{-3}\,{}^{+0.2}_{-0.1}$ & $3.164\times10^{-2}\,{}^{+0.2}_{-0.2}$ \\
    8  & $8.000\times10^{-4}\,{}^{+0.1}_{-0.1}$ & $3.605\times10^{-4}\,{}^{+0.1}_{-0.1}$ & $2.640\times10^{-3}\,{}^{+0.1}_{-0.1}$
       & $1.718\times10^{-3}\,{}^{+0.1}_{-0.1}$ & $7.711\times10^{-4}\,{}^{+0.2}_{-0.1}$ & $5.702\times10^{-3}\,{}^{+0.2}_{-0.1}$ \\
    10 & $1.461\times10^{-4}\,{}^{+0.1}_{-0.1}$ & $6.663\times10^{-5}\,{}^{+0.1}_{-0.1}$ & $4.786\times10^{-4}\,{}^{+0.1}_{-0.1}$
       & $3.737\times10^{-4}\,{}^{+0.1}_{-0.1}$ & $1.689\times10^{-4}\,{}^{+0.1}_{-0.1}$ & $1.229\times10^{-3}\,{}^{+0.1}_{-0.1}$ \\
  \end{tabular}
  \end{ruledtabular}
\end{table*}

\begin{table*}[t]
  \caption{\nlonnll{} cross sections (in fb) for the charged-current
  channel $E^\pm N$ (sum of both charges) and the neutral channel
  $N\bar N$ of the doublet, and for the triplet charged-current channel
  $\Sigma^\pm\Sigma^0$, obtained as twice the $E^\pm N$ cross section
  (Sec.~\ref{sec:sim}). Scale and PDF uncertainties are given as in
  Table~\ref{tab:xsec}.}
  \label{tab:xsec_cc}
  \begin{ruledtabular}
  \begin{tabular}{c ccc ccc}
    & \multicolumn{3}{c}{$\sqs = 84.6\TeV$} & \multicolumn{3}{c}{$\sqs = 100\TeV$} \\
    $M_L$ [TeV] & $E^\pm N$ & $N\bar N$ & $\Sigma^\pm\Sigma^0$
                & $E^\pm N$ & $N\bar N$ & $\Sigma^\pm\Sigma^0$ \\
    \hline
    1  & $50.81^{+1.2}_{-1.1}$ & $14.05^{+1.2}_{-1.1}$ & $101.6^{+1.2}_{-1.1}$
       & $66.07^{+1.4}_{-1.3}$ & $18.37^{+1.4}_{-1.3}$ & $132.1^{+1.4}_{-1.3}$ \\
    2  & $3.629^{+0.6}_{-0.5}$ & $0.9661^{+0.6}_{-0.5}$ & $7.257^{+0.6}_{-0.5}$
       & $5.008^{+0.7}_{-0.6}$ & $1.347^{+0.7}_{-0.6}$ & $10.02^{+0.7}_{-0.6}$ \\
    4  & $0.1675^{+0.3}_{-0.2}$ & $4.246\times10^{-2}\,{}^{+0.3}_{-0.2}$ & $0.3349^{+0.3}_{-0.2}$
       & $0.2633^{+0.3}_{-0.2}$ & $6.758\times10^{-2}\,{}^{+0.3}_{-0.2}$ & $0.5265^{+0.3}_{-0.2}$ \\
    6  & $1.879\times10^{-2}\,{}^{+0.1}_{-0.1}$ & $4.583\times10^{-3}\,{}^{+0.1}_{-0.1}$ & $3.758\times10^{-2}\,{}^{+0.1}_{-0.1}$
       & $3.420\times10^{-2}\,{}^{+0.2}_{-0.1}$ & $8.451\times10^{-3}\,{}^{+0.2}_{-0.1}$ & $6.840\times10^{-2}\,{}^{+0.2}_{-0.1}$ \\
    8  & $2.941\times10^{-3}\,{}^{+0.1}_{-0.1}$ & $6.994\times10^{-4}\,{}^{+0.1}_{-0.1}$ & $5.883\times10^{-3}\,{}^{+0.1}_{-0.1}$
       & $6.276\times10^{-3}\,{}^{+0.1}_{-0.1}$ & $1.516\times10^{-3}\,{}^{+0.1}_{-0.1}$ & $1.255\times10^{-2}\,{}^{+0.1}_{-0.1}$ \\
    10 & $5.373\times10^{-4}\,{}^{+0.1}_{-0.1}$ & $1.262\times10^{-4}\,{}^{+0.1}_{-0.1}$ & $1.075\times10^{-3}\,{}^{+0.1}_{-0.1}$
       & $1.374\times10^{-3}\,{}^{+0.1}_{-0.1}$ & $3.256\times10^{-4}\,{}^{+0.1}_{-0.1}$ & $2.748\times10^{-3}\,{}^{+0.1}_{-0.1}$ \\
  \end{tabular}
  \end{ruledtabular}
\end{table*}

%% ===================================================================
\section{Event simulation}
\label{sec:sim}
%% ===================================================================

Signal events are generated at LO with
MadGraph5\_aMC@NLO and the NN23LO1 PDF set, and are normalised to the \nlonnll{} cross sections of
Sec.~\ref{sec:xsec}, assuming that higher-order corrections do
not change the selection efficiencies. The mixing with the SM lepton is
set to $\epsilon_\ell = 0.01$, which ensures prompt decays, and the
widths of the heavy leptons are fixed to $1\GeV$ in the event generation.
Events are showered and hadronised with Pythia~8, which also decays
the $W$, $Z$ and Higgs bosons in all decay modes. Each production channel and decay
combination is simulated separately and weighted by its branching
fraction.

For the doublet mixing with the tau lepton we generate the three
production channels $E^+E^-$, $E^\pm N$ and $N\bar N$ in all decay modes
($E\to Z\tau, h\tau$ and $N\to W\tau$). Each mode is simulated with
$10^4$ events for $M_L = 2$--$7\TeV$ at $84.6\TeV$, in steps of
$1\TeV$ up to $5\TeV$ and of $0.5\TeV$ above. At $100\TeV$ the
same mass points are used, with 2000 events at $2\TeV$; since the reach at this energy exceeds $7\TeV$, we add samples of 5000 events at 8 and $9\TeV$. For the doublets mixing with electrons or
muons we use 5000 events per decay combination and mass point, with
masses from 2 to $10\TeV$ at $84.6\TeV$ and from 2 to $12\TeV$ at
$100\TeV$, in steps of $1\TeV$ with additional intermediate points at
$84.6\TeV$.

For the singlet mixing with the tau lepton, each decay mode of the pair
($WW$, $WZ$, $Wh$, $ZZ$, $Zh$ and $hh$, with
$\tilde E \to W\nu_\tau, Z\tau, h\tau$) is simulated with 5000 events
for $M_L = 2$--$6\TeV$ ($2$--$7\TeV$ at $100\TeV$) in steps of
$0.5\TeV$. For the singlets mixing with electrons or muons the samples
cover masses from 2 to $8\TeV$ at $84.6\TeV$ and from 2 to $9\TeV$ at
$100\TeV$ in steps of $0.5\TeV$, again with 5000 events per decay
combination and mass point.

For the triplet, pairs are produced through
$q\bar q \to \gamma^*/Z^* \to \Sigma^+\Sigma^-$ and
$q\bar q' \to W^{*\pm} \to \Sigma^\pm\Sigma^0$. The $W\Sigma^\pm\Sigma^0$
coupling is $g$, compared with $g/\sqrt2$ for the $WEN$ coupling of the
doublet, so that $\sigma(\Sigma^\pm\Sigma^0) = 2\,\sigma(E^\pm N)$. We
have verified this relation at LO at $M = 5\TeV$, where the ratio is
$2.008$. Accordingly, we normalise $\Sigma^\pm\Sigma^0$ to twice the
\nlonnll{} $E^\pm N$ cross section (Table~\ref{tab:xsec_cc}) and
$\Sigma^+\Sigma^-$ to the triplet \nlonnll{} cross section of
Table~\ref{tab:xsec}. At $M = 5\TeV$ the charged-current channel
accounts for about two thirds of the total rate. For all three
flavours the triplet samples contain 5000 events per decay combination
and mass point, with masses from 2 to $10\TeV$ at $84.6\TeV$ and from
2 to $12\TeV$ at $100\TeV$, in steps of $1\TeV$ with an additional
point at $7.5\TeV$ for $84.6\TeV$.

For the multilepton signal regions we simulate the prompt backgrounds
$t\bar tZ$, $t\bar tH$,
$t\bar tW$, $WZ$, $ZZ$, $Wh$, $Zh$, $WWW$, $WWZ$, $WZZ$ and $ZZZ$ at LO
with the NN23LO1 PDF set. Unless stated otherwise, the top quarks and
heavy bosons in the background samples are decayed in Pythia~8 in all
decay modes. Since the multilepton signal regions (Sec.~\ref{sec:analysis})
select events with large $\LT$, the scalar sum of the transverse
momenta of the selected leptons and taus, we require
$\sum p_T > 3\TeV$ for the final-state particles of the hard process at
generator level in order to populate this region efficiently. The same requirement is
applied after showering to the visible final-state particles with
$|\eta| < 2.7$, before the detector simulation. The resulting cross sections
and sample sizes are listed in Table~\ref{tab:bkg}. The multiboson
samples have the same size at $84.6$ and $100\TeV$, while for the $t\bar tV$
processes three times more events are generated at $84.6\TeV$. With
these sample sizes, the background estimate in signal regions with very
small backgrounds is not dominated by statistical fluctuations of
individual weighted events.

Since the singlet rarely yields multilepton final states, it is searched
for in events with an opposite-sign lepton pair and a boosted $V/h$ jet
(Sec.~\ref{sec:analysis}). For this search we additionally simulate
Drell--Yan production with one jet
($pp\to\ell^+\ell^- j$) and dileptonic $t\bar t$, $tW$ and $W^+W^-$
production at $84.6$ and $100\TeV$, with the same generator-level $\sum p_T$
requirement (Table~\ref{tab:bkg}). The
Drell--Yan sample, in which the $V/h$-tagged jet originates from a QCD
jet with a large groomed mass, dominates the background in all
opposite-sign signal regions. The signal regions
for the highest masses require $S_T \gtrsim 5\TeV$, where $S_T$ is the
scalar sum of the transverse momenta of the two leptons and the leading
$V/h$ jet (Sec.~\ref{sec:analysis}), and fewer than one thousand events
of the inclusive sample remain in this region. We therefore
generate an additional high-$\sum p_T$ slice of $\ell^+\ell^- j$ with
$\sum p_T > 4.5\TeV$ at generator level (Table~\ref{tab:bkg}). Events
with a reconstructed $S_T > 5\TeV$ are taken from this slice and those
with $S_T < 5\TeV$ from the inclusive sample, which increases the number
of simulated background events in the signal regions near the exclusion
boundary by a factor of 6--10. The LO cross section of the slice is 15--30\% smaller than
that of the inclusive sample in the same kinematic region, a difference
comparable to the spread of about $\pm 20\%$ between independent
integrations of the inclusive cross section; we treat it as an
uncertainty.

For the tau-coupled singlet search (Sec.~\ref{sec:analysis}), whose
signal contains hadronic taus and boosted $V/h$ jets, we separately
simulate the backgrounds with a real or misidentified $\taueh$, listed
in the lowest block of Table~\ref{tab:bkg}. These are
$\tau^+\tau^- j$, $\ell\nu j$ ($\ell = e,\mu,\tau$), the inclusive
dibosons $WW$, $WZ$ and $ZZ$, inclusive $t\bar t$ and $tW$, and QCD
dijets, again generated with $\sum p_T > 3\TeV$. For $\tau^+\tau^- j$, $\ell\nu j$, the
dibosons and the signal, at least one charged lepton ($e$, $\mu$ or
$\tau$) with $p_T > 80\GeV$ is required after the decays. The
$t\bar t$, $tW$ and dijet samples contribute mainly through
misidentified taus and are not filtered.

All three groups of background samples are normalised to their LO
cross sections, listed in Table~\ref{tab:bkg}. Since higher-order QCD
corrections to diboson and $t\bar tV$ production at large transverse
momenta can be sizeable, and since the jet mass in $\ell^+\ell^- j$ is
described by a single parton, we estimate their impact by rescaling
all backgrounds by a common factor $K_B$, whose effect on the reach is
discussed in Sec.~\ref{sec:results:flav}.

\begin{table}[t]
  \caption{Background cross sections at LO after the generator-level
  requirement $\sum p_T > 3\TeV$ ($4.5\TeV$ for the high-$\sum p_T$
  slice of $\ell^+\ell^- j$), and numbers of generated events
  ($N_{\rm gen}$), at $\sqs = 84.6\TeV$ and $100\TeV$. The processes in
  the middle block ($\ell = e,\mu$; both $W$ bosons decay
  leptonically) are used only in the opposite-sign singlet search, and those in the lowest block (all decay
  modes) only in the search for the tau-coupled singlet.}
  \label{tab:bkg}
  \begin{ruledtabular}
  \begin{tabular}{lcccc}
    & \multicolumn{2}{c}{$84.6\TeV$} & \multicolumn{2}{c}{$100\TeV$} \\
    Process & $\sigma$ [fb] & $N_{\rm gen}$ & $\sigma$ [fb] & $N_{\rm gen}$ \\
    \hline
    $t\bar tZ$  & 56.0  & $1.5\times10^6$ & 87.0  & $5\times10^5$ \\
    $t\bar tH$  & 29.3  & $9\times10^5$   & 45.9  & $3\times10^5$ \\
    $t\bar tW$  & 15.4  & $6\times10^5$   & 21.1  & $2\times10^5$ \\
    $WZ$        & 4.03  & $3\times10^5$   & 5.54  & $3\times10^5$ \\
    $ZZ$        & 2.27  & $3\times10^5$   & 3.10  & $3\times10^5$ \\
    $Wh$        & 1.06  & $2\times10^5$   & 1.43  & $2\times10^5$ \\
    $Zh$        & 0.564 & $2\times10^5$   & 0.769 & $2\times10^5$ \\
    $WWW$       & 1.58  & $2\times10^5$   & 2.19  & $2\times10^5$ \\
    $WWZ$       & 3.99  & $3\times10^5$   & 5.54  & $3\times10^5$ \\
    $WZZ$       & 0.675 & $3\times10^5$   & 0.938 & $3\times10^5$ \\
    $ZZZ$       & 0.120 & $10^5$          & 0.167 & $10^5$ \\
    \hline
    $\ell^+\ell^- j$                 & 5.65  & $2\times10^5$   & 8.14  & $2\times10^5$ \\
    \ \ $\sum p_T>4.5\TeV$ & 0.729 & $3\times10^5$ & 1.11 & $2.6\times10^5$ \\
    $t\bar t\to\ell^+\ell^-$ & 55.3 & $8\times10^5$ & 85.4 & $8\times10^5$ \\
    $tW\to\ell^+\ell^-$  & 4.38  & $1.5\times10^5$ & 6.88  & $1.5\times10^5$ \\
    $WW\to\ell^+\ell^-$ & 0.515 & $3\times10^4$  & 0.714 & $3\times10^4$ \\
    \hline
    $\tau^+\tau^- j$                 & 3.14  & $2\times10^5$   & 4.80  & $2\times10^5$ \\
    $\ell\nu j$ ($\ell = e,\mu,\tau$) & 426   & $2\times10^5$   & 596   & $2\times10^5$ \\
    $WW$, $WZ$, $ZZ$                 & 17.1  & $2\times10^5$   & 23.6  & $2\times10^5$ \\
    $t\bar t$                        & 1178  & $2\times10^5$   & 1816  & $2\times10^5$ \\
    $tW$                             & 92.3  & $10^5$          & 146   & $10^5$ \\
    $jj$                             & $3.49\times10^5$ & $3\times10^5$ & $5.23\times10^5$ & $3\times10^5$ \\
  \end{tabular}
  \end{ruledtabular}
\end{table}

The showered signal and background events are then passed through a
fast simulation of the detector response with Delphes~\cite{deFavereau:2013fsa}, using the
July 2022 release of the FCC-hh scenario~II card distributed with
Delphes~3.5. This card is the official parametrisation of the FCC-hh
reference detector of Refs.~\cite{FCC:2018vvp,FCC:2025lpp}. Hadronic taus are
tagged with a flat efficiency of 0.6 and a mistag rate of 0.01 for jets
with $|\eta|<2.5$~\cite{Bhattiprolu:2019vdu}. For the tau-coupled
singlet search, Delphes is run with a truth matching of jets to visible
hadronic tau decays, and the tagging efficiency and mistag rate are
applied as event weights.
Pileup is not simulated, although up to
$\langle\mu\rangle \approx 1000$ simultaneous interactions per bunch
crossing are expected at the FCC-hh design
luminosity~\cite{FCC:2018vvp}. Because our signal regions rely on
leptons and taus with transverse momenta of several hundred GeV and on
values of $\LT$ in the multi-TeV range, the additional pileup
activity, which is mostly soft, has only a small relative
effect on these kinematic variables; its main impact is expected on
lepton isolation and hadronic-$\tau$ identification. The FCC-hh detector design relies on
precision timing and charged-particle vertexing to mitigate these
effects~\cite{FCC:2018vvp}, and we assume that the efficiencies of the
scenario~II card can be maintained.

%% ===================================================================
\section{Analysis strategy}
\label{sec:analysis}
%% ===================================================================

The final states of VLL pair production depend on the representation
and on the lepton flavour. Doublets decay only into $Z\ell$, $h\ell$
and $W\ell$, so that pair production can yield three or more charged
leptons with little SM background. The triplet in addition produces
same-sign lepton pairs, since $\Sigma^0$ is a Majorana fermion. For
both we use multilepton signal regions (SRs) at large $\LT$. Singlets decay
predominantly into $W\nu$ and rarely produce more than two charged
leptons. We therefore search for them in final states with an
opposite-sign lepton or tau pair and a boosted $Z$ or $h$ boson. Below we first
define the common objects and then describe the searches for the
doublet, the singlet and the triplet. The signal regions are summarised
in Table~\ref{tab:sr}.

\begin{table*}[t]
  \caption{Summary of the signal regions. In the multilepton SRs
  ($\ell = e,\mu$) the $\LT$ threshold is chosen for each mass from
  $\{3, 3.5, 4, 4.5, 5, 6, 7\}\TeV$. In the singlet searches the
  window width $w$ of the lepton--jet (tau--jet) mass window
  $[M_L(1-w), M_L(1+w/2)]$ and the $S_T$ threshold are optimised for each
  mass and category, and the 1J and 2J categories contain one and at
  least two $V/h$-tagged jets, respectively.}
  \label{tab:sr}
  \begin{ruledtabular}
  \begin{tabular}{lll}
    Search & Signal region & Requirements \\
    \hline
    \multicolumn{3}{l}{Multilepton (doublet, triplet): $\geq 2\ell$, $p_T(\ell_1) > 200\GeV$, no $b$-tagged jet} \\
    & $2\ell$SS              & two same-sign $\ell$, no $\taueh$ \\
    & $2\ell$SS+$\MET$       & $2\ell$SS, $\MET > 300\GeV$ \\
    & $2\ell$SS$1\taueh$     & two same-sign $\ell$, $\geq 1\,\taueh$ \\
    & $2\ell$SS$1\taueh$+$\MET$ & $2\ell$SS$1\taueh$, $\MET > 150\GeV$ \\
    & $3\ell$-$Z$            & three $\ell$, no $\taueh$, OSSF pair within $15\GeV$ of $m_Z$ \\
    & $3\ell$-no$Z$          & three $\ell$, no $\taueh$, no such pair \\
    & $3\ell$-no$Z$+$\MET$   & $3\ell$-no$Z$, $\MET > 300\GeV$ \\
    & $3\ell 1\taueh$        & $\geq 3\ell$, $\geq 1\,\taueh$ \\
    & $3\ell 1\taueh$-no$Z$  & $3\ell 1\taueh$ with $Z$ veto \\
    & $4\ell$                & $\geq 4\ell$ \\
    & $2\ell 2\taueh$        & $\geq 2\ell$, $\geq 2\,\taueh$ \\
    & $2\ell 2\taueh$-no$Z$  & $2\ell 2\taueh$ with $Z$ veto \\
    \hline
    Singlet ($e$, $\mu$) & 1J, 2J & exactly two opposite-sign $\ell$, $p_T(\ell_2) > 300\GeV$, $m_{\ell\ell} > 200\GeV$; \\
    & & $V/h$ jet with $\tau_{21} < 0.6$; $w \in \{0.10, 0.15, 0.20\}$; $S_T/M_L \in \{0, 0.8, 1.0, 1.2\}$ \\
    \hline
    Singlet ($\tau$) & $\ell\taueh$ (1J, 2J) & one $\ell$ with $p_T > 200\GeV$, $\geq 1$ opposite-sign $\taueh$ \\
    & $\taueh\taueh$ (1J, 2J) & two opposite-sign $\taueh$, no $\ell$ above $100\GeV$ \\
    & & $V/h$ jet with $\tau_{21} < 0.45$; $w \in \{0.15, 0.25, 0.35\}$; \\
    & & $S_T/M_L \in \{0, 0.8, \ldots, 2.4\}$ with $S_T \geq 3\TeV$ ($S_T$ includes $\MET$) \\
  \end{tabular}
  \end{ruledtabular}
\end{table*}

In all searches, electrons and muons must have $p_T > 100\GeV$ and
$|\eta| < 2.5$. In the multilepton SRs, tagged hadronic taus must
satisfy the same requirements, and the leptons must be separated by
$\Delta R > 0.1$ from each other and by $\Delta R > 0.3$ from jets
with $p_T > 30\GeV$.

\begin{figure*}[t]
  \centering
  \includegraphics[width=0.48\textwidth]{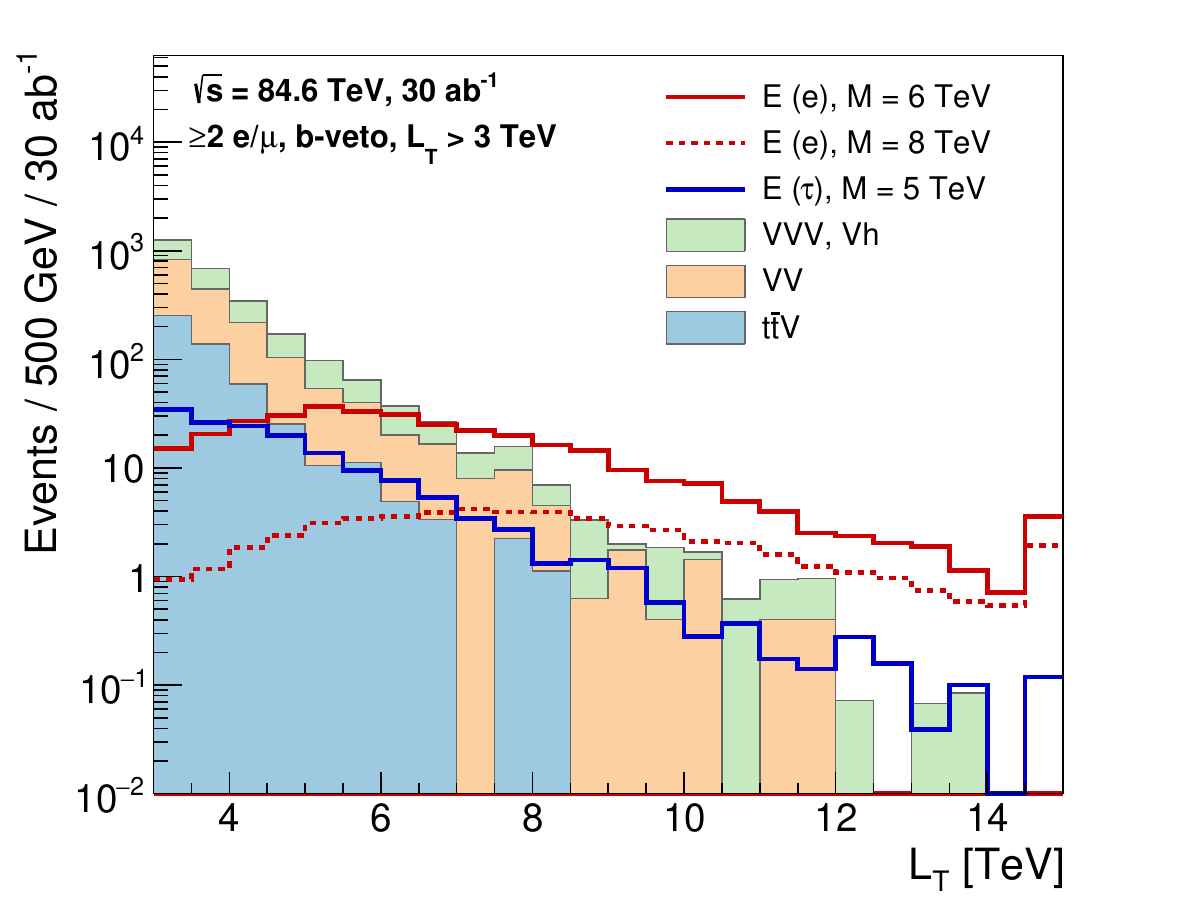}\hfill
  \includegraphics[width=0.48\textwidth]{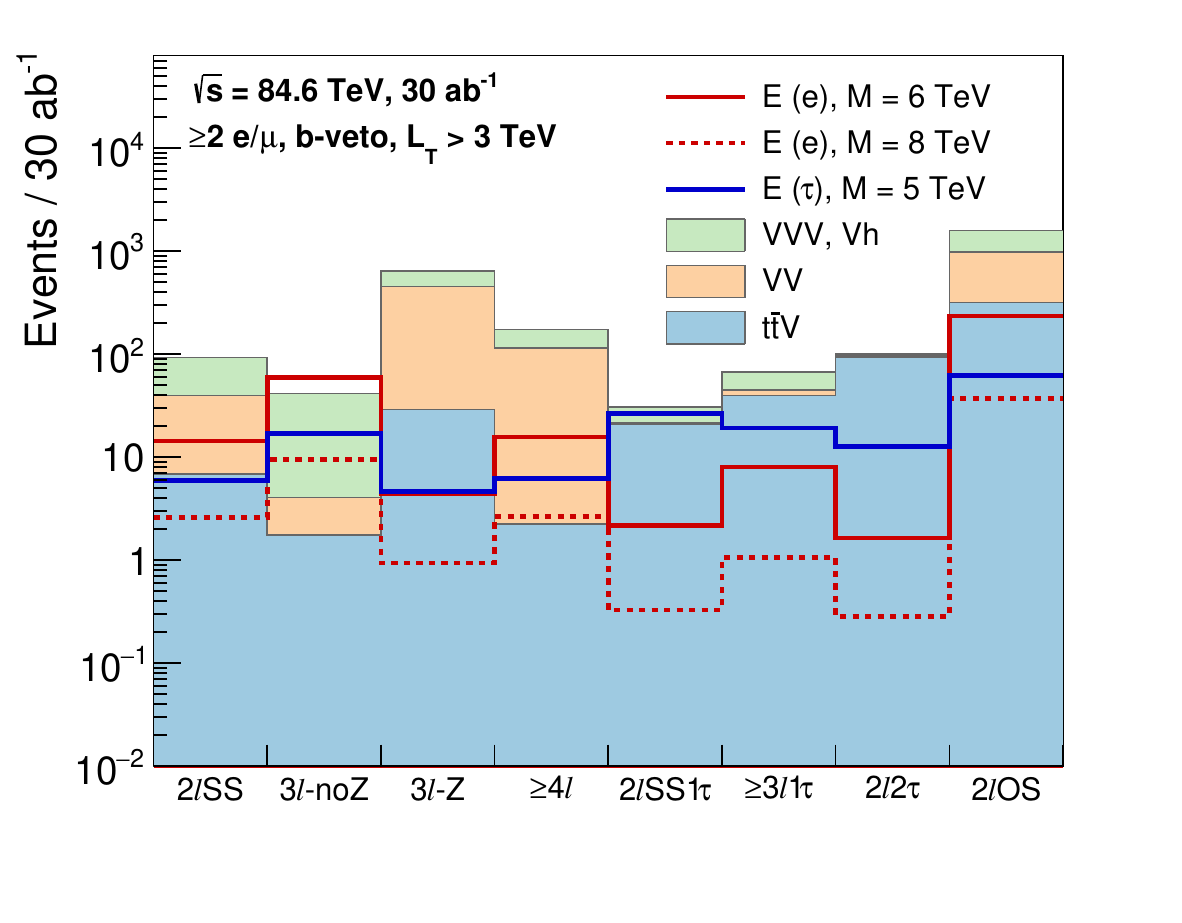}
  \caption{$\LT$ distribution (left) and expected event yields in
  exclusive lepton categories (right) at $\sqs = 84.6\TeV$ with
  $30\abinv$, after the flavour-general preselection ($\geq 2$ light
  leptons, $b$ veto) and $\LT > 3\TeV$. The backgrounds are stacked. The
  signal (\nlonnll{} normalisation) is shown for the electron doublet
  with $M = 6$ and $8\TeV$ and for the tau doublet with $M = 5\TeV$. In
  the left panel the last bin includes the overflow. The $2\ell$OS
  category (two opposite-sign light leptons and fewer than two
  $\taueh$) does not enter the multilepton SRs.}
  \label{fig:kin}
\end{figure*}

For the doublets we require at least two light leptons, a leading
light lepton above $200\GeV$ and no $b$-tagged jet, and on this basis
define twelve SRs. Three of them contain two same-sign light leptons, namely
$2\ell$SS without $\taueh$, $2\ell$SS+$\MET$, which adds
$\MET > 300\GeV$, and $2\ell$SS$1\taueh$ with $\geq 1\,\taueh$. Another
three contain exactly three light leptons without $\taueh$, either with
($3\ell$-$Z$) or without ($3\ell$-no$Z$) an opposite-sign same-flavour
pair within $15\GeV$ of $m_Z$, the latter also in a variant with
$\MET > 300\GeV$. Three further SRs require $\geq 3$ light leptons with
$\geq 1\,\taueh$ ($3\ell 1\taueh$), $\geq 4$ light leptons ($4\ell$),
or $\geq 2$ light leptons with $\geq 2\,\taueh$ ($2\ell 2\taueh$). The
remaining three tighten the tau regions by requiring $\MET > 150\GeV$
in $2\ell$SS$1\taueh$ and the $Z$ veto in $3\ell 1\taueh$ and
$2\ell 2\taueh$. In each SR the $\LT$ threshold is chosen from
$\{3, 3.5, 4, 4.5, 5, 6, 7\}\TeV$, and for each mass we use the most
sensitive combination of SR and $\LT$ threshold, separately for
exclusion and discovery. Figure~\ref{fig:kin} shows the $\LT$
distribution and the event yields in the main lepton categories. The
background falls steeply with $\LT$ and consists mainly of diboson
events (mostly in the $3\ell$-$Z$ category) and $t\bar tV$ events,
whereas the signal extends to $\LT \approx M$ and beyond. For the
light-lepton doublets the $3\ell$-no$Z$ category has the best
signal-to-background ratio. For the tau doublet the signal is spread
over several categories, and those containing a $\taueh$ are the most
sensitive at high $\LT$. Since only a few background events pass the
highest $\LT$ thresholds, we further require $s \geq 3$ for
exclusion.

Singlets rarely produce same-sign or three-lepton final states
(Sec.~\ref{sec:results:flav}), so the singlet searches use events with an opposite-sign lepton pair and a hadronically decaying
$Z$ or $h$. At multi-TeV masses the $Z/h$ decay products merge into a
single anti-$k_T$ $R=0.4$ jet of the FCC-hh card. A jet $J$ is
$V/h$-tagged if it has $p_T > 300\GeV$, a soft-drop mass
$65 < m_{\rm SD} < 145\GeV$ and an N-subjettiness ratio $\tau_{21}$
below $0.45$ in the tau search and below $0.6$ in the light-lepton
search. Events with one (at least two) tagged jets form the exclusive
1J (2J) category. In the 1J category the invariant mass of $J$ and one
of the leptons must lie in the window $[M_L(1-w), M_L(1+w/2)]$. In the
2J category we compare the two lepton--jet pairings and require the mean
mass of the pairing with the smaller mass difference to lie in the
window.

For the singlet mixing with the tau lepton, only the $Z\tau$ and
$h\tau$ decays produce a charged lepton. Two taus are thus present only
in the $ZZ$, $Zh$ and $hh$ channels, which make up a quarter of the
pairs. We therefore search for an opposite-sign tau pair together with
a $V/h$-tagged jet. Hadronic-tau candidates are
jets with $p_T > 200\GeV$, a mass below $50\GeV$ and no $b$ tag,
tagged with the efficiency and mistag rate given in Sec.~\ref{sec:sim}.
No $\Delta R$ separation between leptons, taus and jets is applied in
this search. Of its two channels, the $\ell\taueh$ channel requires exactly one
light lepton ($p_T > 200\GeV$, from a leptonic tau decay) and at least
one tagged $\taueh$ of opposite charge, whereas the $\taueh\taueh$
channel requires two opposite-sign tagged $\taueh$ and no light lepton
above $100\GeV$. The missing transverse momentum is assigned to the tau
candidate closest in azimuth (collinear approximation). Each channel is
divided into the 1J and 2J categories, and in each category the window
width is chosen from $w \in \{0.15, 0.25, 0.35\}$
and the threshold on $S_T = p_T(\tau_1) + p_T(\tau_2) + p_T(J_1) +
\MET$ from $\{0, 0.8, \ldots, 2.4\}\times M_L$ in steps of $0.2$. This
threshold is kept at or above $3\TeV$, the generator-level threshold of
the backgrounds. The four categories are combined in quadrature.

For the singlet mixing with electrons or muons we exploit the decay
chain $\tilde E^+\tilde E^- \to (Z/h\,\ell^+)(Z/h\,\ell^-,\,W\nu)$. We
require exactly two opposite-sign leptons of the benchmark flavour with
$p_T(\ell_2) > 300\GeV$ and $m_{\ell\ell} > 200\GeV$, reject events with
a third lepton, and require at least one $V/h$-tagged jet. In this
search leptons are separated by $\Delta R > 0.3$ only from jets with
$p_T > 150\GeV$ and a soft-drop mass outside $60$--$150\GeV$, so that
leptons close to a $V/h$-jet candidate are retained. The window
width
$w \in \{0.10, 0.15, 0.20\}$ and a threshold on
$S_T = p_T(\ell_1) + p_T(\ell_2) + p_T(J_1)$ of
$\{0, 0.8, 1.0, 1.2\}\times M_L$ are optimised for each mass and
category, and the two categories are then combined in quadrature.

For the triplet we apply the twelve multilepton SRs defined above to all
three flavours without further optimisation.

%% ===================================================================
\section{Results and discussion}
\label{sec:results}
%% ===================================================================

This section presents the projected exclusion and discovery reach of
the searches defined in Sec.~\ref{sec:analysis} for all benchmark
scenarios. In each SR the expected
numbers of signal and background events are
\begin{equation}
  s = \mathcal{L} \sum_i \sigma_i \, \mathcal{B}_i \, \varepsilon_i ,
  \qquad
  b = K_B \, \mathcal{L} \sum_j \sigma_j \, \varepsilon_j ,
\end{equation}
where $\mathcal{L} = 30\abinv$ unless stated otherwise, $\sigma_i$ is the
\nlonnll{} cross section of the production channel (Sec.~\ref{sec:xsec}),
$\mathcal{B}_i$ the branching fraction of the decay combination $i$ and
$\varepsilon_i$ the fraction of its simulated events that pass the
selection, weighted with the tau-tagging probabilities in the
tau-coupled singlet search. The sum over $j$ runs over the background
processes of Table~\ref{tab:bkg}, with LO cross sections $\sigma_j$ and
efficiencies $\varepsilon_j$, and $K_B = 1$ unless stated otherwise.
For a background uncertainty $\Delta b$ we use the median expected
discovery and exclusion significances~\cite{Kumar:2015tna,Cowan:2010js}
\begin{align}
  Z_{\rm disc} &= \Big\{ 2\Big[(s+b)\ln\frac{(s+b)(b+\Delta b^2)}{b^2+(s+b)\Delta b^2}
     \nonumber\\ &\qquad
     - \frac{b^2}{\Delta b^2}\ln\Big(1+\frac{\Delta b^2\, s}{b(b+\Delta b^2)}\Big)\Big]\Big\}^{1/2}, \\
  Z_{\rm excl} &= \Big\{ 2\Big[s - b\ln\frac{b+s+x}{2b}
     - \frac{b^2}{\Delta b^2}\ln\frac{b-s+x}{2b}\Big]
     \nonumber\\ &\qquad
     - (b+s-x)\Big(1+\frac{b}{\Delta b^2}\Big)\Big\}^{1/2},
\end{align}
with $x = [(s+b)^2 - 4sb\,\Delta b^2/(b+\Delta b^2)]^{1/2}$. A mass is excluded
at 95\% CL if $Z_{\rm excl} > 1.645$ and $s \geq 3$, and can be
discovered if
$Z_{\rm disc} > 5$ and $s \geq 10$. In both cases the background is
set to at least one event, and results are shown for
$\Delta b/b = 0.1$ and $0.5$. Because the $s \geq 3$ and $s \geq 10$
requirements make $Z$ discontinuous in mass, for the pure-flavour
benchmarks we interpolate the \nlonnll{} cross section logarithmically
between its tabulated values in $0.5\TeV$ steps and the selection
efficiencies linearly between the simulated masses, and then evaluate
$Z$ on a $25\GeV$ grid.

\subsection{Pure-flavour benchmarks at $84.6\TeV$}
\label{sec:results:flav}
We first present the reach for the nine pure-flavour benchmarks at
$84.6\TeV$, in which a doublet, singlet or triplet VLL mixes
exclusively with $e$, $\mu$ or $\tau$. The results are summarised in
Table~\ref{tab:reachfg}, which also lists the values at $100\TeV$
discussed in Sec.~\ref{sec:results:100}. For the doublet and triplet
the reach is obtained with the multilepton SRs for all three flavours,
whereas for the singlet we use the opposite-sign dilepton search for
electrons and muons and the tau-pair search for the tau-coupled
singlet, to which the multilepton SRs are insensitive. The signal
cross-section uncertainties shift the reach by at most $0.2\TeV$.
Larger, asymmetric uncertainties occur for the tau doublet at
$\Delta b/b = 0.5$ and the tau triplet at $100\TeV$, where a shift of the cross section changes
which $\LT$ threshold satisfies the $s \geq 3$ condition.

\begin{table}[t]
  \caption{Projected 95\% CL exclusion and $5\sigma$ discovery reach
  (in TeV) with $30\abinv$ and \nlonnll{} signal normalisation for the
  pure-flavour benchmarks, obtained with the searches of
  Sec.~\ref{sec:analysis}, at $\sqs = 84.6\TeV$ and $100\TeV$. The
  uncertainties are the combined scale and PDF uncertainties of the
  signal cross section, and the reach is evaluated on a $25\GeV$ grid.
  A missing uncertainty indicates that the corresponding variation of
  the cross section leaves the reach unchanged on this grid, and an
  entry ${<}\,2$ indicates that the reach lies below $2\TeV$, the lowest
  simulated mass.}
  \label{tab:reachfg}
  \begin{ruledtabular}
  \begin{tabular}{lcccc}
    & \multicolumn{2}{c}{$\Delta b/b = 0.1$} & \multicolumn{2}{c}{$\Delta b/b = 0.5$} \\
    Benchmark & Excl. & Disc. & Excl. & Disc. \\
    \hline
    \multicolumn{5}{l}{$\sqs = 84.6\TeV$} \\
    Doublet $\tau$  & $6.40^{+0.13}_{-0.18}$ & $4.78^{+0.10}_{-0.08}$ & $6.40^{+0.13}_{-0.33}$ & $4.05^{+0.10}_{-0.08}$ \\
    Doublet $e$     & $9.30^{+0.18}_{-0.13}$ & $7.83^{+0.10}_{-0.08}$ & $9.28^{+0.15}_{-0.13}$ & $7.45^{+0.13}_{-0.13}$ \\
    Doublet $\mu$   & $9.65^{+0.15}_{-0.10}$ & $8.23^{+0.13}_{-0.10}$ & $9.63^{+0.15}_{-0.13}$ & $7.85^{+0.13}_{-0.13}$ \\
    Singlet $\tau$  & $3.45^{+0.05}_{-0.03}$ & $<2$ & $2.43^{+0.03}$ & $<2$ \\
    Singlet $e$     & $5.68^{+0.08}_{-0.10}$ & $4.28^{+0.03}_{-0.03}$ & $5.63^{+0.08}_{-0.18}$ & $2.13$ \\
    Singlet $\mu$   & $6.03^{+0.08}_{-0.05}$ & $4.45^{+0.10}_{-0.05}$ & $5.95^{+0.10}_{-0.05}$ & $2.13_{-0.03}$ \\
    Triplet $\tau$  & $5.58^{+0.10}_{-0.08}$ & $3.45^{+0.05}_{-0.08}$ & $5.45^{+0.10}_{-0.05}$ & $2.90^{+0.05}_{-0.05}$ \\
    Triplet $e$     & $8.38^{+0.15}_{-0.13}$ & $6.83^{+0.13}_{-0.10}$ & $8.33^{+0.15}_{-0.13}$ & $6.28^{+0.13}_{-0.08}$ \\
    Triplet $\mu$   & $8.75^{+0.18}_{-0.13}$ & $7.20^{+0.10}_{-0.08}$ & $8.70^{+0.18}_{-0.10}$ & $6.75^{+0.13}_{-0.10}$ \\
    \hline
    \multicolumn{5}{l}{$\sqs = 100\TeV$} \\
    Doublet $\tau$  & $7.80^{+0.13}_{-0.10}$ & $5.88^{+0.15}_{-0.10}$ & $7.80^{+0.13}_{-0.10}$ & $5.45^{+0.13}_{-0.05}$ \\
    Doublet $e$     & $10.45^{+0.15}_{-0.13}$ & $8.70^{+0.13}_{-0.13}$ & $10.45^{+0.15}_{-0.13}$ & $8.23^{+0.13}_{-0.10}$ \\
    Doublet $\mu$   & $10.93^{+0.15}_{-0.13}$ & $9.18^{+0.15}_{-0.10}$ & $10.93^{+0.15}_{-0.13}$ & $8.73^{+0.15}_{-0.10}$ \\
    Singlet $\tau$  & $3.60^{+0.10}_{-0.05}$ & $<2$ & $<2$ & $<2$ \\
    Singlet $e$     & $6.33^{+0.05}_{-0.08}$ & $4.65^{+0.05}_{-0.05}$ & $6.23^{+0.05}_{-0.08}$ & $2.23$ \\
    Singlet $\mu$   & $6.73^{+0.05}_{-0.10}$ & $4.90^{+0.03}_{-0.10}$ & $6.43^{+0.15}_{-0.05}$ & $2.30_{-0.05}$ \\
    Triplet $\tau$  & $6.28^{+0.15}_{-0.08}$ & $4.30^{+0.05}_{-0.08}$ & $6.15^{+0.28}_{-0.08}$ & $4.03^{+0.05}_{-0.08}$ \\
    Triplet $e$     & $9.60^{+0.18}_{-0.13}$ & $7.55^{+0.18}_{-0.13}$ & $9.60^{+0.18}_{-0.13}$ & $6.90^{+0.15}_{-0.10}$ \\
    Triplet $\mu$   & $10.13^{+0.20}_{-0.15}$ & $7.93^{+0.18}_{-0.10}$ & $10.13^{+0.18}_{-0.15}$ & $7.43^{+0.13}_{-0.10}$ \\
  \end{tabular}
  \end{ruledtabular}
\end{table}

For the tau-coupled doublet the exclusion (discovery) reach is
$6.4\TeV$ ($4.78\TeV$) for $\Delta b/b = 0.1$, and the sensitivity
comes mainly from the $2\ell 2\taueh$ and $3\ell 1\taueh$ regions with
tight $\LT$ thresholds. For doublets mixing with light leptons the reach is driven by the
$3\ell$-no$Z$ region at $\LT > 6$--$7\TeV$, which is essentially free
of background (about one expected event at $\LT > 6\TeV$). The exclusion reach is $9.3\TeV$ for electrons and
$9.65\TeV$ for muons and depends only weakly on $\Delta b/b$, while the
discovery reach is $7.8\TeV$ and $8.2\TeV$, respectively. The somewhat larger reach for muons
reflects the higher muon efficiency of the FCC-hh card at multi-TeV
transverse momenta. The theoretical uncertainty of the signal cross
section (scale $\oplus$ PDF) translates into an uncertainty of
${}^{+0.15}_{-0.10}\TeV$ to ${}^{+0.18}_{-0.13}\TeV$ on the light-lepton
reach. Doubling the background ($K_B = 2$) at
$\Delta b/b = 0.1$ lowers the exclusion reach by only $75\GeV$ for the
electron and muon doublets and leaves it unchanged for the tau doublet,
whereas the discovery reach decreases by $150$--$175\GeV$ and
$0.5\TeV$, respectively.

An integrated luminosity of $30\abinv$ remains the target of the FCC-hh
programme at the new baseline energy~\cite{Blondel:2025kbl}. To
illustrate the dependence on luminosity, we have repeated the analysis
for $10$ and $3\abinv$. For $\Delta b/b = 0.1$ the exclusion reach for
the tau-coupled doublet decreases from $6.40\TeV$ to $5.53\TeV$ at
$10\abinv$ and to $4.58\TeV$ at $3\abinv$, and that for the electron
(muon) doublet from $9.30\TeV$ ($9.65\TeV$) to $7.95\TeV$ ($8.35\TeV$)
and $6.70\TeV$ ($7.08\TeV$). Each factor of three in luminosity thus
changes the exclusion reach by about $0.9\TeV$ for $\tau$ and by about
$1.3\TeV$ for $e$ and $\mu$.

The singlet is much harder to probe in multilepton final states. Its
dominant decay, $\tilde E \to W\nu$, yields at most one charged lepton,
while the decays $\tilde E \to Z\ell, h\ell$ mostly lead to two
opposite-sign leptons accompanied by hadronic $Z/h$ decays. Only
$\sim 0.1$--$0.2\%$ of the singlet events enter the same-sign and
$\geq 3\ell$ regions, which limits the multilepton exclusion reach to
$2.43\TeV$ ($e$) and $2.68\TeV$ ($\mu$) at $\Delta b/b = 0.1$ and
leaves no discovery sensitivity.

The tau-coupled singlet is probed with the tau-pair search of
Sec.~\ref{sec:analysis}. The background is dominated by
$\ell\nu j$ events with a misidentified $\taueh$ and $V/h$ jet in the
$\ell\taueh$ channel and by QCD dijets in the $\taueh\taueh$ channel,
followed by $\tau^+\tau^- j$. At $M_L = 3.5\TeV$, for example, the four
categories together contain $s = 19$ signal events (1.6\% of the
produced pairs) and $b = 162$ background events, of which $s = 9$ and
$b = 93$ fall in the $\ell\taueh$ 1J category.

The exclusion reach is $3.45\TeV$ at $\Delta b/b = 0.1$ and $2.43\TeV$
at $\Delta b/b = 0.5$, where the latter value is poorly defined because
the expected exclusion significance stays close to the 95\% CL
threshold over a wide mass range. A $5\sigma$ discovery is not possible above
$2\TeV$, the lowest simulated mass (Table~\ref{tab:reachfg}). Owing to
the small signal-to-background ratio, the result depends on the
misidentification rate and on the background normalisation. For a
mistag rate of 0.005 (0.02) the exclusion reach is $3.70\TeV$
($2.78\TeV$), and with $K_B = 2$ it is $2.93\TeV$. With $10\abinv$ the
exclusion reach is $2.68\TeV$. A search for one hadronic tau, two $V/h$
jets and large $\MET$ could extend the reach by targeting the $WZ$ and
$Wh$ channels, but would require a dedicated simulation of
$Z(\nu\bar\nu)+$jets and larger top and diboson samples.

For the singlet coupling to electrons or muons, the opposite-sign
dilepton search of Sec.~\ref{sec:analysis} selects about 10\% of the
singlet events. The background consists almost
entirely of Drell--Yan $\ell^+\ell^- j$ events, with the dileptonic $t\bar t$, $tW$ and $WW$
backgrounds and the multiboson backgrounds of Table~\ref{tab:bkg}
contributing only at the per cent level. At $M_L = 4\TeV$,
for example, the 1J (2J) category of the muon benchmark contains
$s = 39$ ($24$) signal and $b = 140$ ($7.3$) background events for
$30\abinv$.

At $\Delta b/b = 0.1$ the resulting exclusion reach
(Table~\ref{tab:reachfg}) is $5.7\TeV$ for electrons and $6.0\TeV$ for
muons, more than twice the multilepton result, and the discovery reach
is $4.3\TeV$ and $4.45\TeV$, respectively. Above $5\TeV$ the 2J category contains fewer than ten signal events, so that the discovery
reach is set by the 1J category
and depends more strongly on the background level than the exclusion
reach. For $\Delta b/b = 0.5$ the exclusion reach is $5.6\TeV$ ($e$)
and $5.95\TeV$ ($\mu$), whereas the discovery reach drops to $2.1\TeV$.
Doubling the background ($K_B = 2$) lowers the exclusion reach by at
most $0.2\TeV$ and the discovery reach by $0.35$--$0.85\TeV$ at
$\Delta b/b = 0.1$. Scaling the high-$\sum p_T$ slice of
$\ell^+\ell^- j$ up by one third (Sec.~\ref{sec:sim}) lowers the
exclusion reach by $0.03$--$0.18\TeV$. With $10\abinv$ ($3\abinv$) the exclusion reach is
$4.65$--$5.0\TeV$ ($3.6$--$3.9\TeV$). At $3\abinv$ this exceeds the
reach of about $2.5\TeV$ estimated for a $100\TeV$ collider by
extrapolating LHC multilepton searches~\cite{Guedes:2021oqx}.

\begin{figure*}[t]
  \centering
  \includegraphics[width=\textwidth]{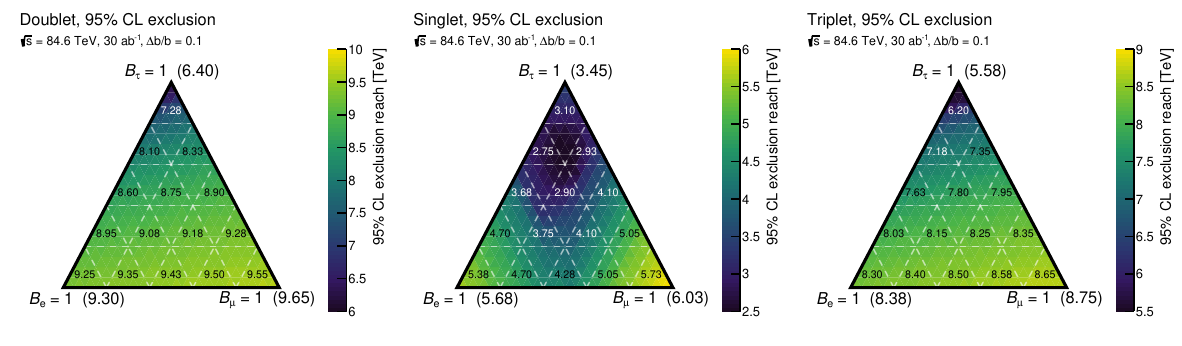}
  \caption{95\% CL exclusion reach for the doublet (left), singlet
  (middle) and triplet (right) VLL in the plane of the flavour branching
  fractions, $\mathcal{B}_e + \mathcal{B}_\mu + \mathcal{B}_\tau = 1$,
  at $\sqs = 84.6\TeV$ with $30\abinv$ and $\Delta b/b = 0.1$. The
  numbers give the reach in TeV at the centres of the sub-triangles,
  and the values in parentheses are the pure-flavour results of
  Table~\ref{tab:reachfg}. For the doublet and triplet the efficiency
  for mixed-flavour pairs is approximated by the mean of the
  pure-flavour efficiencies; for the singlet only same-flavour pairs
  are counted (see text). The colour scales differ between the panels.}
  \label{fig:triangle}
\end{figure*}

The triplet is an intermediate case. As for the
singlet, half of the $\Sigma^\pm$ decays go to $W\nu$, but every $\Sigma^0 \to W\ell$ decay
yields a charged lepton. Since $\Sigma^0$ is a Majorana fermion, about one third of the dilepton events
in the $\Sigma^\pm\Sigma^0 \to (Z/h\,\ell)(W\ell)$ channels are
same-sign. For the tau-coupled triplet the SRs
with hadronic taus ($3\ell 1\taueh$, $2\ell 2\taueh$ and
$2\ell$SS$1\taueh$+$\MET$) give an exclusion reach of $5.58\TeV$ and a
discovery reach of $3.45\TeV$, $0.8$--$1.3\TeV$ below those of the
tau-coupled doublet. For $\ell = e$
($\mu$) the exclusion reach is $8.38\TeV$ ($8.75\TeV$) and the discovery
reach $6.83\TeV$ ($7.20\TeV$) for $\Delta b/b = 0.1$. The sensitivity
comes mainly from the $3\ell$-no$Z$ region at $\LT > 4$--$6\TeV$ and,
close to the exclusion limit, from the $2\ell$SS region. This reach
lies about $0.9\TeV$ below that of the doublet, since only half of the
$\Sigma^\pm$ decays produce a charged lepton, but well above that of
the singlet in the same final states. Doubling the background lowers the triplet
exclusion reach by $175$--$250\GeV$ ($e$, $\mu$) and $125\GeV$ ($\tau$)
and the discovery reach by $250$--$325\GeV$ and $375\GeV$, somewhat
more than for the doublet. For all multilepton benchmarks the exclusion
reach thus changes by at most 3\%.
With $10\abinv$ ($3\abinv$) the triplet exclusion reach at
$\Delta b/b = 0.1$ decreases to $7.2$--$7.6\TeV$ ($6.1$--$6.45\TeV$)
for $e$ and $\mu$ and to $4.45\TeV$ ($3.4\TeV$) for $\tau$. For $e$
and $\mu$ this corresponds to about $1.15\TeV$ per factor of three in
luminosity, slightly less than for the light-lepton doublets, and for
$\tau$ to about $1.05$--$1.1\TeV$.

\subsection{General flavour mixing}
\label{sec:results:mix}
For general mixing, a VLL decays into a lepton of flavour $\ell$,
summed over the $W$, $Z$ and $h$ channels, with the flavour branching
fraction $\mathcal{B}_\ell = |\epsilon_\ell|^2 / \sum_{\ell'}
|\epsilon_{\ell'}|^2$, while the boson branching fractions of
Sec.~\ref{sec:models} remain unchanged. A pair therefore decays into the
flavours $(\ell,\ell')$ with probability
$\mathcal{B}_\ell \mathcal{B}_{\ell'}$. We have simulated only
pure-flavour pairs. For the doublet and the triplet, which are analysed
with the multilepton SRs for all flavours, we approximate the
efficiency of a mixed pair in each signal region and $\LT$ bin by the
mean of the two pure-flavour efficiencies. The expected signal is then
linear in the branching fractions, $s = \sum_\ell \mathcal{B}_\ell\,
s_{\ell\ell}$, and the optimal signal region and $\LT$ threshold are
chosen at each point. For the tau-coupled doublet, whose samples extend
to $7\TeV$, we keep the efficiencies constant above this mass. The
singlet searches are flavour specific, since the opposite-sign dilepton
search requires exactly two electrons or two muons and the tau search
requires a tau pair. We therefore count only same-flavour pairs, whose signal scales
with $\mathcal{B}_\ell^2$, and combine the $ee$, $\mu\mu$ and
$\tau\tau$ channels, which are mutually exclusive, in quadrature.
Mixed pairs, which would require for instance an $e\mu$ category, are neglected, so that the singlet result is conservative away from the
corners.

Figure~\ref{fig:triangle} shows the resulting 95\% CL
exclusion reach in the plane of the flavour branching fractions for
the three models. For the doublet and the triplet it decreases smoothly
from the light-lepton corners towards the tau corner. For equal
branching fractions, $\mathcal{B}_e = \mathcal{B}_\mu =
\mathcal{B}_\tau = 1/3$, the doublet exclusion (discovery) reach is
$9.0\TeV$ ($7.6\TeV$), and it stays above $8.3\TeV$ ($6.8\TeV$) as
long as $\mathcal{B}_\tau \leq 0.6$. For $\Delta b/b = 0.5$ the
exclusion reach is almost unchanged ($9.0\TeV$ at the centre of the
triangle), whereas the discovery reach falls to $7.1\TeV$ at the centre
and to $4.1\TeV$ at the tau corner. For the triplet the exclusion
(discovery) reach at the centre is $8.1\TeV$ ($6.5\TeV$), and the
exclusion reach stays above $7.35\TeV$ for $\mathcal{B}_\tau \leq
0.6$. The reach for light leptons is thus largely independent of how
the mixing is shared between electrons and muons, and only a sizeable
tau component brings it down towards the pure-tau value. For the
singlet, the quadratic dependence on the branching fractions makes the reach fall towards the interior of the triangle, from about $4.3$--$4.4\TeV$ near the $e$--$\mu$ edge
($\mathcal{B}_e \approx \mathcal{B}_\mu \approx 1/2$) to
$3.55\TeV$ at the centre and about $2.5\TeV$ for $\mathcal{B}_\tau \approx 0.6$--$0.7$, below
the pure-tau value of $3.45\TeV$.

\subsection{Results at $100\TeV$}
\label{sec:results:100}
We have repeated all analyses at $\sqs = 100\TeV$ with the same signal
regions and the signal and background samples of Sec.~\ref{sec:sim};
the results are listed in the lower part of Table~\ref{tab:reachfg}. For $\Delta b/b = 0.1$ the doublet
exclusion reach is $7.80\TeV$ ($\tau$), $10.45\TeV$ ($e$) and
$10.93\TeV$ ($\mu$), and the discovery reach is $5.88$, $8.70$ and
$9.18\TeV$, respectively. The lower baseline energy thus reduces the
light-lepton reach by 10--12\% and that of the tau doublet by about
18\%.

\begin{figure*}[t]
  \centering
  \includegraphics[width=\textwidth]{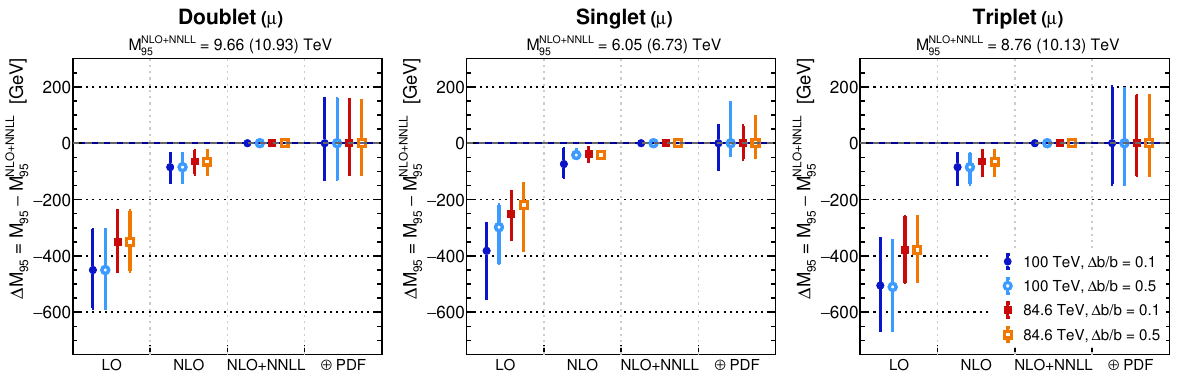}
  \caption{Shift $\Delta M_{95} = M_{95} - M_{95}^{\rm NLO+NNLL}$ of the
  95\% CL exclusion reach with $30\abinv$ for LO, NLO and \nlonnll{}
  signal normalisation, for the muon-coupled doublet (left), singlet
  (middle) and triplet (right), at $\sqs = 84.6\TeV$ and $100\TeV$ and
  for two values of the background uncertainty, using the analyses of
  Table~\ref{tab:reachfg}. The \nlonnll{} reach at $\Delta b/b = 0.1$ is
  given above each panel. The error bars show the scale uncertainty, and
  the last column the combined scale and PDF uncertainty.}
  \label{fig:reach}
\end{figure*}

For the tau-coupled singlet the exclusion reach at $\Delta b/b = 0.1$
increases by only $0.15\TeV$ to $3.60\TeV$. For $\Delta b/b = 0.5$ the
reach is poorly defined, as at $84.6\TeV$, and we therefore quote it
only as below $2\TeV$. This search remains limited by the
backgrounds with misidentified hadronic taus, mainly from $W$ plus jet
and multijet production, which grow with the collision energy as well.
The opposite-sign dilepton plus $Z/h$-jet search excludes singlets
mixing with electrons (muons) up to $6.33\TeV$ ($6.73\TeV$) and
discovers them up to $4.65\TeV$ ($4.90\TeV$), 9--12\% above the
values at $84.6\TeV$. For the triplet, the exclusion (discovery) reach
at $\Delta b/b = 0.1$ is $6.28\TeV$ ($4.30\TeV$) for $\tau$,
$9.60\TeV$ ($7.55\TeV$) for $e$ and $10.13\TeV$ ($7.93\TeV$) for
$\mu$. For light leptons the exclusion reach is 15--16\% above the
values at $84.6\TeV$ and again $0.8$--$0.9\TeV$ below that of the
doublet.

In the flavour triangles at $100\TeV$, the doublet exclusion
(discovery) reach for equal branching fractions is $10.1\TeV$
($8.4\TeV$), and it stays above $9.3\TeV$ ($7.5\TeV$) as long as
$\mathcal{B}_\tau \leq 0.6$. For the triplet the corresponding values
are $9.2\TeV$ ($7.1\TeV$) at the centre and above $8.3\TeV$
($6.1\TeV$) for $\mathcal{B}_\tau \leq 0.6$. For the singlet the
exclusion reach is $3.7\TeV$ at the centre and falls to about
$2.4\TeV$ for $\mathcal{B}_\tau \approx 0.6$, as at $84.6\TeV$.

\subsection{Impact of precision predictions}
\label{sec:results:prec}
Figure~\ref{fig:reach} shows how the 95\% CL exclusion reach depends on
the perturbative order of the signal normalisation, with the selection
efficiencies kept fixed. To compare the three representations under the
same conditions, we use the muon-coupled benchmarks with the analyses of
Sec.~\ref{sec:results:flav} (Table~\ref{tab:reachfg}) and show the
shift
\begin{equation}
  \Delta M_{95} = M_{95} - M_{95}^{\rm NLO+NNLL} ,
\end{equation}
where $M_{95}$ is the 95\% CL exclusion reach obtained with the signal
normalisation at a given perturbative order and $M_{95}^{\rm NLO+NNLL}$
the reach with the central \nlonnll{} normalisation.
In this section the reach is evaluated on a finer mass grid ($5\GeV$,
or $1\GeV$ for the singlet) than the $25\GeV$ grid of
Table~\ref{tab:reachfg}, so that the \nlonnll{} values can differ from
those in the table by up to $20\GeV$. The NLO and \nlonnll{} cross
sections are those of Sec.~\ref{sec:xsec}. For the singlet and triplet
the relative LO scale band is taken from the doublet $E^+E^-$ channel,
since the relative factorisation-scale dependence at LO agrees to within
$0.2\%$ between the $E^+E^-$, $E^\pm N$ and $N\bar N$ channels, which
involve different quark couplings.

The markers in Fig.~\ref{fig:reach} show how the central reach changes
with the perturbative order. For $\Delta b/b = 0.1$, replacing the LO
normalisation (CT18NNLO,
$\mu_0 = M_{LL}$) by NLO increases the doublet exclusion reach at
$84.6\TeV$ from $9.31\TeV$ to $9.60\TeV$, and \nlonnll{} raises it
further to $9.66\TeV$. At $100\TeV$ the corresponding values are
$10.48 \to 10.85 \to 10.93\TeV$. The singlet reach increases from $5.80$
to $6.05\TeV$ at $84.6\TeV$ and from $6.35$ to $6.73\TeV$ at
$100\TeV$, and the triplet reach from $8.38$ to $8.76\TeV$ and from
$9.63$ to $10.13\TeV$. Going from LO to \nlonnll{} thus shifts the
exclusion reach by $+0.25$--$0.5\TeV$, or 4--6\%, for all three
representations. This is consistent with the nearly
representation-independent ratio
$\sigma_{\rm NLO+NNLL}/\sigma_{\rm LO} \simeq 1.26$--$1.35$ at these
masses, which increases with the mass, so that the shift in GeV is smallest for the singlet, whose reach is lowest.

The error bars in Fig.~\ref{fig:reach} show the theoretical uncertainty
of the reach at each order. For $\Delta b/b = 0.1$ the scale uncertainty
amounts to about
$\pm 80$--$170\GeV$ at LO, $\pm 25$--$60\GeV$ at NLO and at most
$\pm 5\GeV$ at \nlonnll. At \nlonnll{} the uncertainty is thus
dominated by the PDFs, which give
${}^{+155}_{-110}$ to ${}^{+160}_{-130}\GeV$ for the doublet,
${}^{+60}_{-55}$ to ${}^{+60}_{-95}\GeV$ for the singlet and
${}^{+170}_{-115}$ to ${}^{+200}_{-145}\GeV$ for the triplet. A further
improvement therefore requires better constraints on the large-$x$
quark and antiquark densities.

The $5\sigma$ discovery reach behaves similarly. For the doublet it
increases from $7.89$ to $8.23\TeV$ at $84.6\TeV$ and from $8.82$ to
$9.20\TeV$ at $100\TeV$, and for the triplet from $6.89$ to $7.21\TeV$
and from $7.54$ to $7.94\TeV$, an increase of $0.32$--$0.40\TeV$. The
scale uncertainty at \nlonnll{} again stays below $5\GeV$.

The other flavours follow the same pattern. For the tau-coupled
doublet and triplet, whose reach is lower, the shift from LO to
\nlonnll{} at $\Delta b/b = 0.1$ is $+0.32$--$0.37\TeV$ for the
triplet and $+0.48\TeV$ ($+0.41\TeV$) at $84.6\TeV$ ($100\TeV$) for the
doublet. The larger shift for the tau doublet arises because the
optimised $\LT$ thresholds of the multilepton SRs become tighter as the
signal rate increases, so that the reach depends more strongly on the
cross section (cf.\ the asymmetric uncertainties in
Table~\ref{tab:reachfg}).

%% ===================================================================
\section{Complementarity with other facilities}
\label{sec:compl}
%% ===================================================================
\begin{figure}[t]
  \centering
  \includegraphics[width=\columnwidth]{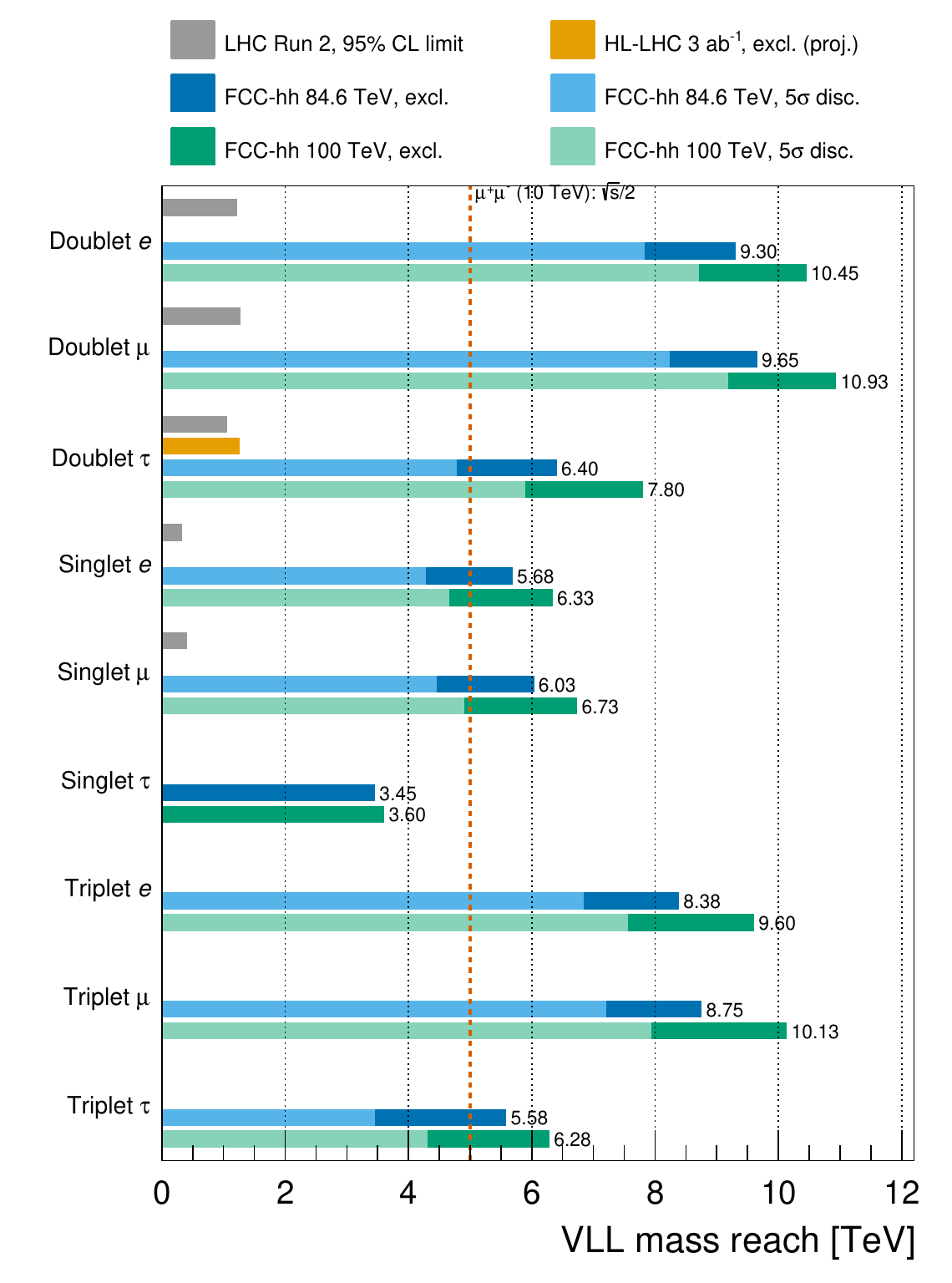}
  \caption{Mass reach for doublet, singlet and triplet VLLs coupling to
  $e$, $\mu$ or $\tau$. Shown are the LHC Run~2 lower limits at 95\% CL
  (ATLAS~\cite{ATLAS:2024mrr} for $e$ and $\mu$, CMS~\cite{CMS:2022nty}
  for $\tau$), the HL-LHC projection of Ref.~\cite{Bhattiprolu:2019vdu}
  for the tau-coupled doublet, and the FCC-hh exclusion and discovery
  reach obtained in this work at $84.6\TeV$ and $100\TeV$ ($30\abinv$,
  $\Delta b/b = 0.1$, Table~\ref{tab:reachfg}). No $5\sigma$ discovery is
  possible for the tau-coupled singlet. The LHC limits on the triplet,
  $0.85$--$1.07\TeV$ depending on the flavour
  structure~\cite{CMS:2022nty,ATLAS:2022yhd}, are not available per
  flavour and are therefore not shown. The dashed line marks the
  kinematic limit $\sqs/2$ of a $10\TeV$ muon collider.}
  \label{fig:compl}
\end{figure}

Whether FCC-hh offers a unique window on VLLs depends on how its reach
compares with that of other colliders. Hadron colliders probe the mass
directly through Drell--Yan pair production, whereas lepton colliders
constrain the mixing with the SM leptons through precision
measurements, and a high-energy muon collider can do both up to its
kinematic limit. In this section we first compare the FCC-hh results
obtained in this work at $84.6\TeV$ and $100\TeV$ with the current LHC
limits and the projections for its upgrades (Fig.~\ref{fig:compl}) and
then discuss FCC-ee and a $10\TeV$ muon collider.

At the LHC, the Run~2 searches exclude doublet VLLs
up to $1.045\TeV$ ($\tau$), $1.22\TeV$ ($e$) and $1.27\TeV$ ($\mu$),
whereas singlets are excluded only in a narrow mass window for $\tau$
and up to $0.32\TeV$ ($e$) and $0.40\TeV$ ($\mu$)~\cite{ATLAS:2024mrr,CMS:2022nty,ATLAS:2023sbu}.
Type-III triplets are excluded up to $0.85$--$1.07\TeV$, depending on
their branching fractions to the three lepton
flavours~\cite{CMS:2022nty}, and up to $0.91\TeV$ for equal branching
fractions~\cite{ATLAS:2022yhd}. For VLLs mixing with the tau
lepton, the projections of Ref.~\cite{Bhattiprolu:2019vdu} indicate
that the HL-LHC with $3\abinv$ could exclude the doublet up to about
$1.25\TeV$ and discover it up to about $0.9\TeV$, a $27\TeV$ HE-LHC
with $15\abinv$ up to about $2.3\TeV$ and $1.7\TeV$, and a $100\TeV$
collider with $30\abinv$ up to about $5.75\TeV$ and $4.0\TeV$; the
minimal singlet remains out of reach at all of these machines. At
$100\TeV$, by contrast, our analysis excludes the tau-coupled doublet up to
$7.80\TeV$ and allows its discovery up to $5.88\TeV$, and it excludes
the minimal singlet up to $3.60\TeV$.
Relative to the current LHC limits, FCC-hh at $84.6\TeV$ improves the
doublet reach by a factor of about six (tau) to almost eight ($e$,
$\mu$), and the triplet reach by a factor of five to ten.

At an $e^+e^-$ collider, VLL pairs can be produced
only up to $M_L = \sqs/2 \le 182.5\GeV$, well below the LHC bounds.
Direct searches at $\sqs = 240\GeV$ have been studied for other VLL
scenarios, for example a VLL accompanying a scalar dark-matter
candidate~\cite{Elgammal:2026ggs}. For the minimal models considered
here, FCC-ee is sensitive only indirectly, since at tree level the
mixing modifies the $Z$ and $W$ couplings of the SM leptons by terms of order
$|\epsilon_\ell|^2$, independently of $M_L$. The $Z$-pole and
$WW$-threshold runs will therefore improve the electroweak bounds on
$\epsilon_\ell$ quoted in Sec.~\ref{sec:models}, but they do not
constrain the mass. Loop effects in Higgs-strahlung are sensitive to
vector-like Yukawa couplings in extended models, but in the
mixing-driven scenarios the existing electroweak bounds already exclude an
observable effect~\cite{Marzo:2026ykm}. Indirect measurements at
FCC-ee and direct searches at FCC-hh thus probe different parameters,
$\epsilon_\ell$ and $M_L$, respectively.

A $10\TeV$ muon collider can pair-produce
VLLs of all multiplets and flavours through electroweak $s$-channel
production in a clean environment, up to the kinematic limit
$M_L = 5\TeV$. We are not aware of a dedicated study of the
reach close to this limit. Above it, a muon collider probes VLLs only
indirectly. For a singlet mixing with the muon, integrating out the
heavy lepton shifts the left-handed $Z$ coupling of the muon by an
amount proportional to the squared mixing angle~\cite{Biswas:2026mab}.
Measurements of $\mu^+\mu^- \to Zh$ and related Higgs production
processes at a $10\TeV$ muon collider would constrain this shift,
together with other muon--Higgs--gauge interactions, up to an order of
magnitude better than current bounds~\cite{Biswas:2026mab}. As at
FCC-ee, this indirect probe constrains the mixing $\epsilon_\mu$ rather
than the mass $M_L$.

At $84.6\TeV$, the FCC-hh exclusion (discovery) reach of $6.4\TeV$
($4.8\TeV$) for the tau-coupled doublet lies close to the kinematic
limit of the muon collider, and the reach of $9.3$--$9.65\TeV$ for the
doublet with $e$ or $\mu$ mixing lies well beyond it. The tau-coupled
singlet, with an FCC-hh exclusion reach of $3.45\TeV$, lies within the
kinematic reach of the muon collider. For
the singlet coupling to $e$ or $\mu$, the opposite-sign dilepton search
at FCC-hh excludes masses up to $5.7$--$6.0\TeV$, slightly above the
muon-collider limit, but its discovery reach of $4.3$--$4.45\TeV$ lies
below it. For the triplet, the FCC-hh exclusion reach of $5.6\TeV$ in
the tau-coupled case is again close to the muon-collider limit, and
that of $8.4$--$8.75\TeV$ for coupling to $e$ or $\mu$ exceeds it.

%% ===================================================================
\section{Conclusions}
\label{sec:concl}
%% ===================================================================

We have presented \nlonnll{} predictions for Drell--Yan pair production
of vector-like leptons at FCC-hh, both at the new baseline energy of $84.6\TeV$ and
at $100\TeV$. Threshold resummation increases the NLO cross section by
3--6\% for masses between 1 and $10\TeV$ and reduces the scale
uncertainty to below 1\% for $M_L \geq 2\TeV$. The remaining theoretical uncertainty comes from the PDFs and grows
from 3--4\% at $3\TeV$ to about 7\% at $7\TeV$ and 10\% at $10\TeV$.
The
transverse momentum of the pair is strongly affected by the parton
shower; with MC@NLO-$\Delta$ matching, NLO+PS exceeds fixed-order NLO by
a factor $1.55$--$1.8$ at $p_T \approx 0.2$--$1\,M_L$, whereas the
integrated rate is unchanged.

In a detector-level multilepton analysis with $30\abinv$ at $84.6\TeV$,
the tau-coupled doublet can be excluded up to $6.4\TeV$ and discovered
up to $4.8\TeV$, and doublets mixing with electrons or muons up to
$9.3$--$9.65\TeV$ and $7.8$--$8.2\TeV$. At $100\TeV$ the reach
increases to $7.8\TeV$ and $5.9\TeV$ for taus, and to
$10.45$--$10.9\TeV$ and $8.7$--$9.2\TeV$ for electrons and muons.

Singlets rarely produce same-sign or three-lepton final states, and
the tau-coupled singlet is not accessible in multilepton final states
at all. A dedicated search for an opposite-sign tau pair and a boosted
$Z/h$ jet excludes it up to $3.45\TeV$, but the large background from
misidentified hadronic taus precludes a discovery. An analogous search
for two opposite-sign electrons or muons and a boosted $Z/h$ jet, whose
background is dominated by Drell--Yan plus jet production, reaches
$5.7$--$6.0\TeV$ (exclusion) and $4.3$--$4.45\TeV$ (discovery) for
singlets coupling to electrons or muons, compared with
$2.4$--$2.7\TeV$ in multilepton final states. At $100\TeV$ the singlet
reach becomes $3.6\TeV$ (exclusion) for taus and $6.3$--$6.7\TeV$
(exclusion) and $4.65$--$4.9\TeV$ (discovery) for electrons and
muons.

Type-III triplets are produced mainly through the charged-current
channel $\Sigma^\pm\Sigma^0$, whose rate is twice that of the doublet
$E^\pm N$ channel, and their Majorana neutral component gives rise to
same-sign dileptons. The multilepton analysis reaches $5.6\TeV$ (exclusion) and
$3.45\TeV$ (discovery) for taus, and $8.4$--$8.75\TeV$ and
$6.8$--$7.2\TeV$ for electrons and muons. For electrons and muons
these values lie about $0.9\TeV$ below those of the doublet, because half of the
$\Sigma^\pm$ decays produce no charged lepton. At $100\TeV$ the triplet
reach becomes $6.3\TeV$ and $4.3\TeV$ for taus, and $9.6$--$10.1\TeV$
and $7.55$--$7.9\TeV$ for electrons and muons. For general flavour
mixing, the doublet (triplet) exclusion reach for equal branching
fractions is $9.0\TeV$ ($8.1\TeV$) at $84.6\TeV$ and stays above
$8.3\TeV$ ($7.35\TeV$) as long as $\mathcal{B}_\tau \leq 0.6$. The
singlet reach, in contrast, falls to $3.55\TeV$ at the centre of the
triangle, because only same-flavour pairs are selected.

Compared with LO, the precision predictions raise the exclusion reach
of the muon-coupled doublet, singlet and triplet by $0.25$--$0.5\TeV$
(4--6\%), almost independently of the representation, and that of the
tau-coupled doublet by $0.4$--$0.5\TeV$. They reduce the scale
uncertainty from $\pm 80$--$170\GeV$ to below $5\GeV$, leaving a PDF
uncertainty of about $60$--$200\GeV$.

Relative to $100\TeV$, the lower baseline energy reduces the exclusion
reach at $30\abinv$ by about 18\% ($1.4\TeV$) for the tau-coupled
doublet, by 11--12\% ($1.15$--$1.3\TeV$) for doublets mixing with
electrons or muons, by about 10\% for singlets mixing with electrons or
muons and by 11--14\% for the triplet. Since a factor of
three in integrated luminosity changes the reach by a similar amount,
the loss is equivalent to roughly a threefold reduction in luminosity.
Already with $3\abinv$, the tau-coupled doublet can be excluded up to
$4.6\TeV$ and doublets mixing with electrons or muons up to
$6.7$--$7.1\TeV$, compared with the current LHC limits of $1.05$--$1.3\TeV$.
Singlets coupling to electrons or muons can be excluded up to
$3.6$--$3.9\TeV$ and triplets coupling to electrons or muons up to
$6.1$--$6.45\TeV$.

\begin{figure*}[t]
  \centering
  \includegraphics[width=0.48\textwidth]{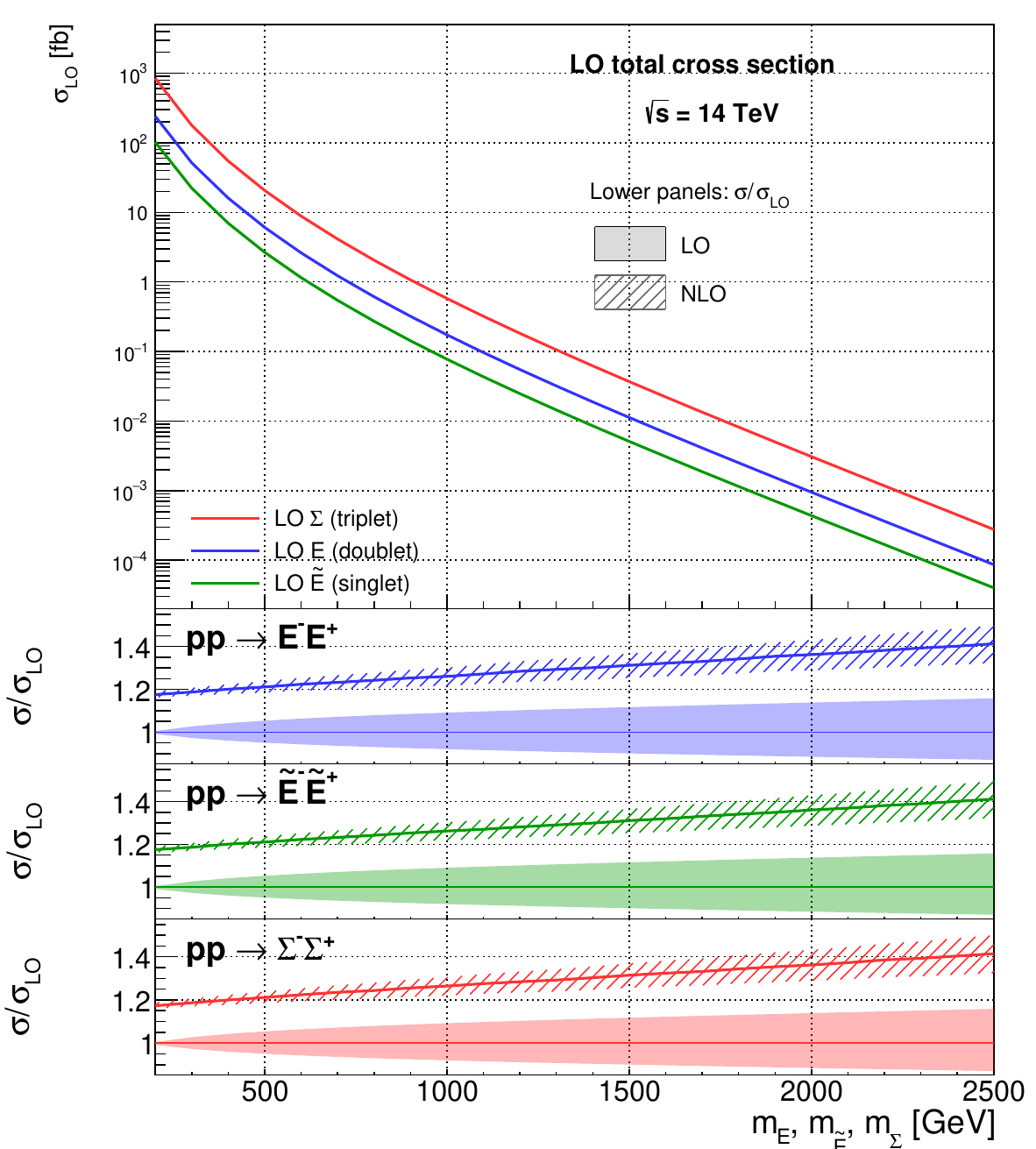}\hfill
  \includegraphics[width=0.48\textwidth]{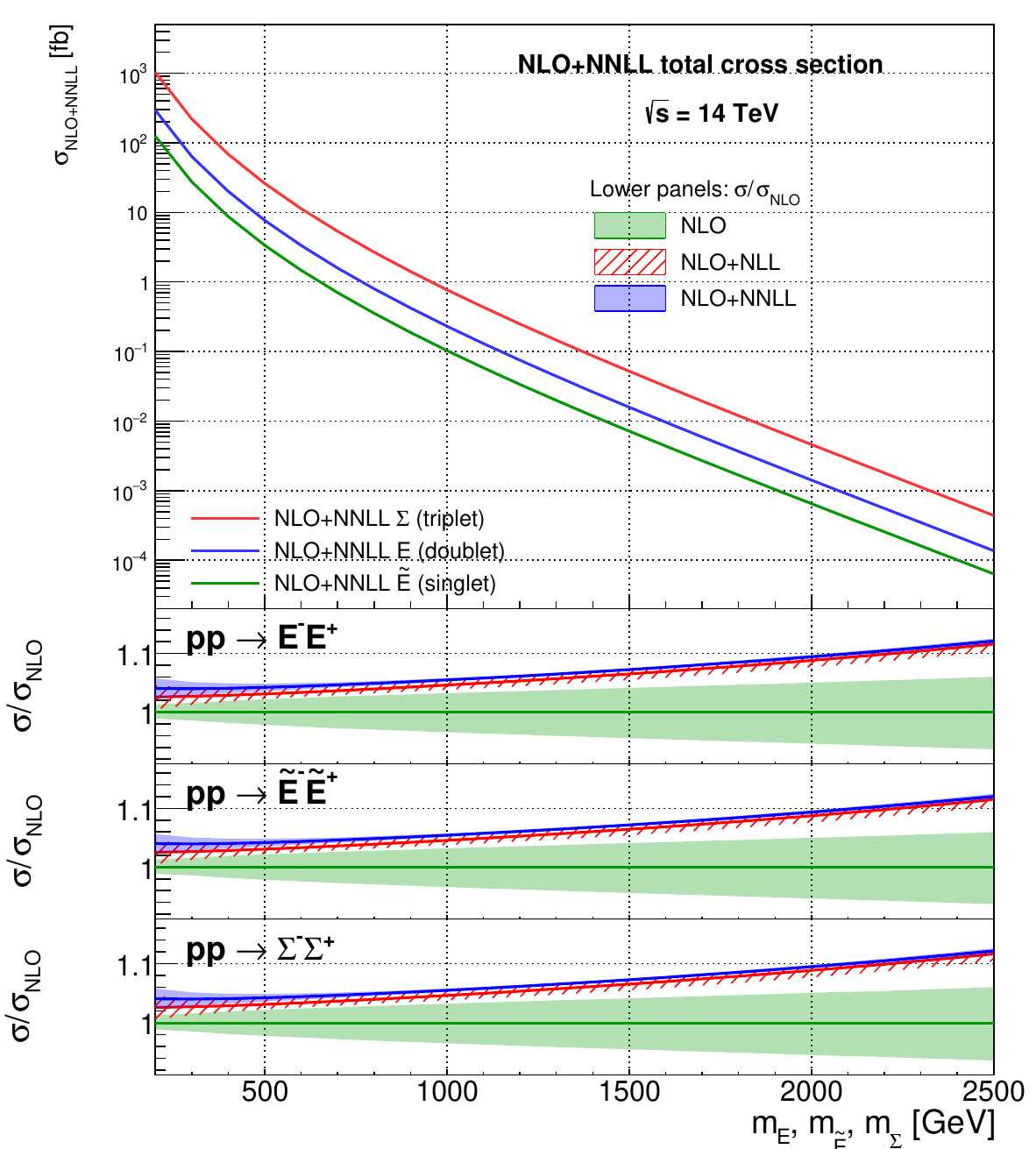}
  \caption{Reproduction of the LHC results of Ref.~\cite{AH:2023hft}
  at $\sqs = 14\TeV$: LO and NLO total cross sections with scale bands
  (left), and \nlonnll{} cross sections with their ratio to NLO
  (right).}
  \label{fig:lhc}
\end{figure*}

With \nlonnll{} predictions, the theoretical uncertainty of the signal is
dominated by PDFs at large Bjorken-$x$ and typically corresponds to $0.1$--$0.2\TeV$ in mass reach (up to $0.3\TeV$ where the $s \geq 3$
requirement shifts the optimal signal region). This is smaller than the
effect of the background systematics on the discovery reach and can be reduced further by PDF
constraints from the HL-LHC, the LHeC or FCC-eh. The reach further
depends on the detector performance at multi-TeV transverse momenta,
beyond the parametrisation used here. Particularly relevant are the momentum
resolution and charge identification of electrons and muons above
several TeV, hadronic-$\tau$ identification, and the tagging of boosted $Z$, $W$ and
$h$ bosons whose decay products merge into a single narrow jet.
A direct search at FCC-hh would cover VLL masses up to about $10\TeV$,
largely independently of the size of the mixing as long as the decays
are prompt, and would complement the indirect sensitivity of FCC-ee and a muon
collider to the mixing through precision measurements. Multivariate
methods, such as mass-parametrised boosted decision trees that exploit
the full event kinematics, could improve on the cut-based selections
used here; we leave such studies to future work.

\begin{acknowledgments}
This work was supported by the Turkish Energy, Nuclear and Mineral
Research Agency (TENMAK) under Grant
No.~2025TENMAK(CERN)A5.H3.F2-05. The numerical calculations reported in
this paper were performed on the server of the High Energy Physics
Laboratory of Bolu Abant Izzet Baysal University (BAIBU), funded by the
Scientific and Technological Research Council of T\"urkiye (TUBITAK)
under Grant No.~114F138.
\end{acknowledgments}

\section*{Data Availability}
The event samples, the analysis
code and the numerical results underlying the figures are available
from the corresponding author upon reasonable request.

%% ===================================================================
\appendix
\section{Validation at the LHC}
\label{app:lhc}
%% ===================================================================

To validate our implementation, we have reproduced the main results of
Ref.~\cite{AH:2023hft} at $\sqs = 14\TeV$ with the setup described in
Sec.~\ref{sec:xsec}. For the total cross sections at LO and NLO
(Fig.~\ref{fig:lhc}, left), the $K$ factor increases from $1.17$ at
$200\GeV$ to $1.41$ at $2.5\TeV$ for all three multiplets, and the NLO
scale uncertainty reaches about $\pm 6\%$ at $2.5\TeV$. Our Mellin-space
LO cross section agrees with MadGraph5\_aMC@NLO to within $0.2\%$.

At \nlonnll{} (Fig.~\ref{fig:lhc}, right) we obtain
\nlonnll/NLO $= 1.040$, $1.055$ and $1.121$ at $0.2$, $1$ and
$2.5\TeV$, with a scale uncertainty below 1\% above $500\GeV$. The LO
and NLO cross sections and these ratios agree with Figs.~1 and~2 of
Ref.~\cite{AH:2023hft} within the precision to which the published
figures can be read.

For the invariant-mass distributions, the integral of our NLO
$d\sigma/dM_{LL}$ reproduces the MadGraph5\_aMC@NLO total to $0.02\%$
for the doublet at $M = 600\GeV$, and the resummed distributions agree
with the original ones; for instance, NLO/\nlonnll{} ranges from $0.96$
to $0.925$ for $M_E = 1.5\TeV$.

We also reproduce the NLO, LO+PS and NLO+PS pair transverse-momentum
distributions for $M = 0.6$, $0.9$ and $1.5\TeV$ with the
MC@NLO-$\Delta$ setup of Sec.~\ref{sec:xsec}. The maximum of
NLO+PS/NLO lies between $1.68$ and $1.99$, compared with about
$1.8$--$2.0$ in Ref.~\cite{AH:2023hft}.

\bibliography{refs}

\end{document}